\documentclass[twocolumn]{aastex61}
\pdfoutput=1

\usepackage{longtable}
\usepackage{graphicx,rotating}
\usepackage{natbib}
\newcommand\aastex{AAS\TeX}

\received{August 3, 2017}
\revised{November 3, 2017}
\accepted{\today}

\shorttitle{\aastex\ Accretion signatures in magnetic HAeBes}
\shortauthors{Reiter et al.}

\begin{document}

\title{Linking signatures of accretion with magnetic field measurements -- line profiles are not significantly different in magnetic and non-magnetic Herbig Ae/Be stars}

\correspondingauthor{Megan Reiter}
\email{mreiter@umich.edu}

\author{Megan Reiter}
\affil{University of Michigan 
311 West Hall, 1085 S. University Ave  
Ann Arbor, MI 48109-1107, USA}

\author{Nuria Calvet}
\affil{University of Michigan 
311 West Hall, 1085 S. University Ave  
Ann Arbor, MI 48109-1107, USA}

\author{Thanawuth Thanathibodee}
\affil{University of Michigan 
311 West Hall, 1085 S. University Ave  
Ann Arbor, MI 48109-1107, USA}

\author{Stefan Kraus}
\affiliation{School of Physics, Astrophysics Group, University of Exeter, Stocker Road, Exeter EX4 4QL, UK}
%
%
\author{P. Wilson Cauley}
\affiliation{Arizona State University}

\author{John Monnier}
\affil{University of Michigan 
311 West Hall, 1085 S. University Ave  
Ann Arbor, MI 48109-1107, USA}

\author{Adam Rubinstein}
\affil{University of Michigan 
311 West Hall, 1085 S. University Ave 
Ann Arbor, MI 48109-1107, USA}

\author{Alicia Aarnio}
\affil{University of Colorado}

\author{Tim J. Harries}
\affiliation{School of Physics, Astrophysics Group, University of Exeter, Stocker Road, Exeter EX4 4QL, UK}



\begin{abstract}
Herbig Ae/Be stars are young, pre-main-sequence stars that sample the transition in structure and evolution between low- and high-mass stars, providing a key test of accretion processes in higher-mass stars. 
Few Herbig Ae/Be stars have detected magnetic fields, calling into question whether the magnetospheric accretion paradigm developed for low-mass stars can be scaled to higher masses. 
We present He~{\sc i}~10830~\AA\ line profiles for 64 Herbig Ae/Be stars with a magnetic field measurement in order to test magnetospheric accretion in the physical regime where its efficacy remains uncertain. 
Of the 5 stars with a magnetic field detection, 1 shows redshifted absorption, indicative of infall, and 2 show blueshifted absorption, tracing mass outflow. 
The fraction of redshifted and blueshifted absorption profiles in the non-magnetic Herbig Ae/Be stars is remarkably similar, suggesting that the stellar magnetic field does not affect gas kinematics traced by He~{\sc i}~10830~\AA. 
Line profile morphology does not correlate with the luminosity, rotation rate, mass accretion rate, or disk inclination. 
Only the detection of a magnetic field and a nearly face-on disk inclination show a correlation (albeit for few sources). 
This provides further evidence for weaker dipoles and more complex field topologies as stars develop a radiative envelope. 
The small number of magnetic Herbig Ae/Be stars has already called into question whether magnetospheric accretion can be scaled to higher masses; accretion signatures are not substantially different in magnetic Herbig Ae/Be stars, casting further doubt that they accrete in the same manner as classical T Tauri stars. 
\end{abstract}

\keywords{stars: formation --- others...}



\section{Introduction} \label{s:intro}

Herbig Ae/Be stars (HAeBes) are the intermediate-mass analogs of low-mass 
T Tauri stars -- young stars that are sufficiently evolved to be studied in 
the optical, but with IR colors that suggest remnant circumstellar material 
and spectroscopic signatures of accretion \citep{her60,hil92,wat98}. 
With masses between $\sim 2-10$~M$_{\odot}$, HAeBes sample 
conditions of formation and evolution intermediate between low- and high-mass stars.
Their stellar structure, multiplicity, pre-main-sequence (PMS) evolution times, and magnetic properties provide insight into the physical parameters that guide the formation of higher-mass stars.
At the same time, they are more numerous than the highest-mass stars and evolve more slowly, allowing them to be studied in the optical and near-IR with techniques similar to those used for T~Tauri stars \citep[e.g.,][]{men12}.

Extensive studies of nearby, low-mass PMS stars led to the development of a magnetically-controlled accretion paradigm \citep[e.g.,][]{cal92,hartmann94,muz98,muz01,muz04,hartmann16}. 
In this picture a strong, predominately dipolar stellar magnetic field interacts with and truncates the circumstellar disk at a few stellar radii.
Magnetic field lines loft material from the disk, guiding it toward the stellar surface where it splashes down at high latitudes. 
Models coevolved with observations to explain 
(1) high-velocity wings in absorption lines \citep[e.g.,][]{edw94,bou99,muz98}, 
(2) excess continuum emission attributed to the accretion shock at the stellar surface \citep[e.g.,][]{cal98,muz98,gul00,muz01}, and 
(3) the detection of strong magnetic fields on T~Tauri stars \citep[e.g.,][]{joh99a,joh07,joh13}.

Changes in the stellar structure above $\gtrsim 1$ M$_{\odot}$ make it unclear 
whether intermediate-mass stars can generate magnetic fields of sufficient 
strength to support magnetospheric accretion like their low-mass counterparts. 
In particular, stars above $\sim 1$~M$_{\odot}$ develop radiative envelopes before they disperse their disks, so they may not maintain strong, ordered magnetic fields throughout their evolution \citep[e.g.,][]{hus09,gre12}.
Weaker fields may not be able to disrupt the disk, leading to smaller infall velocities or possibly a different accretion pathway altogether. 
Surveys of HAeBes find a low magnetic incidence, with fields detected in $\lesssim 10$\% of sources measured \citep[see, e.g.,][]{wad07,ale13,hub13,bag15}. 
To make matters worse, derived upper limits on the magnetic field strength are 
smaller than the minimum field strength required for magnetospheric accretion 
in both Herbig Ae and Be stars \citep[as derived from][]{joh99b}.

Despite the paucity of strongly magnetic HAeBes, observations suggest that
a smooth scaling of the disk-mediated accretion models for low-mass T~Tauri stars
\citep[e.g.,][]{cal04,muz04} can be applied to A-type stars (up to $\sim 4$ M$_{\odot}$). 
It is less clear that magnetospheric accretion models can be applied to higher-mass
B-type stars. 
Recent spectroscopic studies testing whether empirical correlations 
between emission line luminosities and more direct tracers of the accretion 
rate extend to intermediate-mass stars find conflicting results. 
\citet{don11} find a break in the correlation between A- and B-type stars while both 
\citet{men11} and \citet{fai15} find that fitting stellar 
parameters to each star individually produces a correlation that extends 
smoothly to B-type stars.

The line profiles themselves point to a different accretion geometry for earlier spectral types,
if not a different physical mechanism altogether.
\citet{cau14,cau15} report a smaller fraction of HAeBes showing redshifted absorption, indicative of accretion, or blueshifted absorption, tracing outflows, compared to classical T~Tauri stars. 
Maximum velocities in the line profiles are smaller than free-fall, suggesting a more compact accretion geometry. 
\citet{muz04} argued that magnetospheric accretion could proceed for A-type stars through small magnetospheres, since the faster rotation rates of higher-mass stars force corotation to smaller radii. 
Fewer redshifted absorption profiles in the Herbig Be stars compared to the Herbig Aes led \citet{cau14,cau15} to suggest that B-type stars may accrete via a boundary layer \citep[e.g.,][]{ber88,bas89,pop93} rather than through the magnetosphere.

H$\alpha$ spectropolarimetry also hints at different disk geometries around A- and B-type stars \citep{vin02,vin05,mot07,aba17}.
Line polarization is consistent with a gap in the inner disk of Herbig Ae stars, 
as expected if the magnetic field truncates the inner disk at a few stellar radii. 
Unlike the Herbig Aes, observations cannot rule out disks around B-type stars 
that extend to the stellar surface, permitting direct accretion onto the pre-main-sequence star.

Together, these observations illustrate a number of impediments to applying the standard magnetospheric accretion paradigm to HAeBes.
Comparing the magnetic properties to the accretion behavior provides a key test of magnetospheric accretion. 
Whereas magnetospheric accretion has been examined in the context of the stellar magnetic fields measured in T~Tauri stars \citep[e.g.,][]{sym05,joh07}, no such comparison has been made for a large sample of HAeBes.
Fortunately, recent work by \citet{ale13} provides measurements and upper limits on the longitudinal magnetic fields of 70 HAeBes. 
In this paper, we present a combination of new and archival spectra of 64 HAeBes targeted for a magnetic field measurement (63/70 of the \citet{ale13} sample plus one source from the literature) to compare the line profiles of sources with and without evidence for a magnetic field.

Detailed magnetospheric accretion models can reproduce many of the observed features of the Balmer line profiles \citep[e.g.,][]{cal92,hartmann94,cal98,muz98b,gul00,muz01,kur06}.
Among the results of these models is the general prediction that thermalized lines are less likely to show redshifted absorption. 
Line luminosity correlations have been used to argue for a scaling of magnetospheric accretion \citep[e.g.,][]{muz04,men11}, although it is unclear whether accretion dominates the line luminosity in HAeBes \citep[e.g.,][]{men15,fai17}. 
Kinematics provide a more direct assessment of the motion of the gas, and thus a better diagnostic of accretion.

We examine He~{\sc i}~10830~\AA\ line profiles for redshifted or blueshifted absorption profiles that provide direct or indirect evidence of accretion. 
Resonant scattering of permitted lines excited near the young star can produce absorption profiles that trace the kinematics of the gas.
Blueshifted absorption is produced by colder gas moving toward the observer in a wind/outflow from an accreting young star \citep[e.g.,][]{fin84}. 
Redshifted absorption occurs when colder material moves toward the star, and thus away from the observer.
Redshifted absorption can \textit{only} be produced by infall \citep[e.g.,][]{wal72}.

Several papers have explored He~{\sc i}~10830~\AA\ as a tracer of the structure and kinematics of accretion and outflow in young stars \citep[e.g.,][]{tak02,edw03,edw06,fis08,kur11,cau14}.
The $2s^3S$ lower level of the transition is 20~eV above ground, but metastable and therefore long-lived.
Around PMS stars, densities do not tend to be high enough for collisional deexcitation and the lower level is radiatively isolated from the ground state \citep{kwa07}.
The large optical depth, high emissivity, and metastable lower level of the 
line make it particularly susceptible to absorption \citep[see, e.g.,][]{kwa11}, thereby making it a useful tracer of gas kinematics over a range 
of physical conditions.

In this paper, we examine the line profiles of He~{\sc i}~10830~\AA\ for 
evidence of accretion in HAeBes that \citet{ale13} targeted for 
magnetic field measurements. 
We present profiles for 64 HAeBes using a combination of new near-IR spectroscopy from the Folded-port InfraRed Echellette (FIRE) spectrograph on Magellan, X-Shooter (VLT) data from the archive, and GNIRS/PHOENIX spectra from \citet{cau14}. 
We compare the line profiles of the HAeBes with and without a magnetic field detection confirmed by \citet{ale13}.
Including disk inclinations estimated from near-IR H-band long-baseline interferometry by \citet{laz17}, where available, we compare the line profiles with the physical properties of the stars.
Together, these data allow us to probe the role that magnetic fields play in accretion onto HAeBes. 

\section{Observations} \label{s:data}

\subsection{New FIRE spectra}\label{ss:fire}

We present new, near-IR spectra of 33 HAeBes obtained with the Folded-port InfraRed Echellette (FIRE) spectrograph \citep{sim13} on the 6.5~m Baade/Magellan telescope. 
A single FIRE spectrum covers 0.8-2.5~\micron\ with spectral resolution of $R \sim 6000$. 
Data were obtained on three separate nights. 
Objects observed on 2011 March 10 or 12 used a $0\farcs45$ wide slit ($R=8000$) while those observed on 2016 September 20 used a $0\farcs75$ wide slit ($R=4800$). 
Targets were observed using the standard ABBA sequence except for a few bright sources where only two nods were used. 
Wavelength calibration was done using a ThAr lamp. 
Spectra were reduced using the \textsc{firehose} IDL pipeline which performs flat-fielding, object extraction, and flux and wavelength calibration. 
Details for each source are listed in Table~\ref{t:obs}. 


\begin{footnotesize}
\begin{center}
\begin{longtable*}{llllllll}
\caption{He~{\sc i}~10830~\AA\ Spectra}\label{t:obs} \\ 
\hline\hline
HD number & Alt. Name & $\alpha_{\mathrm{J2000}}$ & $\delta_{\mathrm{J2000}}$ & 
Spectrograph & Date & Exp. time (s) \\ 
\endfirsthead 
\hline 
BD-06 1259   & BF Ori    & 05:37:13.3 & --06:35:01 & PHOENIX & 2013 Mar 2  & 800 \\
BD-06 1253   & V380 Ori  & 05:36:25.4 & --06:42:58 & FIRE & 2011 Mar 12 & 3 $\times 4$ \\ 
BD-05 1329   & T Ori     & 05:35:50.4 & --05:28:35 & PHOENIX & 2013 Mar 01 & 800 \\
BD-05 1324   & NV Ori    & 05:35:31.4 & --05:33:08 & FIRE & 2016 Sept 20 & 1 $\times 4$ \\ 
BD+41 3731   &           & 20:24:15.7 & +42:18:01 & PHOENIX & 2013 Nov 12 & 600 \\ 
BD+46 3471   & V1578 Cyg & 21:52:34.1 & +47:13:44 & PHOENIX & 2013 Nov 11 & 600 \\
BD+61 154    & V594 Cas  & 00:43:18.3 & +61:54:40 & PHOENIX & 2013 Mar 3  & 800 \\
HD 17081     & $\pi$~Cet & 02:44:07.3 & --13:51:31 & PHOENIX & 2013 Mar 3 & 90 \\ 
HD~31293     & AB~Aur    & 04:55:45.8 & +30:33:04 & FIRE & 2011 Mar 10 & 5 $\times 4$ \\ 
HD~31648     & MWC 480   & 04:58:46.3 & +29:50:37 & FIRE & 2011 Mar 12 & 3 $\times 4$ \\ 
HD 34282     & V1366 Ori & 05:16:00.5 & --09:48:35 & PHOENIX & 2013 Mar 01 & 800 \\
HD 35187     &           & 05:24:01.2 & +24:57:38 & GNIRS & 2012 Dec 14 & 110 \\
HD 35929     &           & 05:27:42.8 & --08:19:38 & X-shooter & 2009 Dec 17 & 5 $\times 4$ \\
HD 36112     & MWC 758   & 05:30:27.5 & +25:19:57 & FIRE & 2011 Mar 12 & 3 $\times 4$ \\ 
HD~36910     & CQ Tau    & 05:35:58.5 & +24:44:54 & PHOENIX & 2013 Feb 27 & 600 \\
HD 36917     & V372 Ori  & 05:34:46.9 & --05:34:14 & FIRE & 2016 Sept 20 & 1 $\times 4$ \\ 
HD 36982     & LP Ori    & 05:35:09.8 & --05:27:53 & FIRE & 2016 Sept 20 & 1 $\times 4$ \\ 
HD 37258     & V586 Ori  & 05:36:59.1 & --06:09:18 & X-shooter & 2010 Jan 02 & 10 $\times 4$ \\
HD 37357     &           & 05:37:47.1 & --06:42:30 & X-shooter & 2010 Feb 05 & 10 $\times 4$ \\
HD~37806     & MWC 120   & 05:41:02.3 & --02:43:01 & FIRE & 2011 Mar 10 & 1 $\times 4$ \\ 
HD 38120     &           & 05:43:11.9 & --04:59:49 & FIRE & 2016 Sept 20 & 2 $\times 4$ \\ 
HD 38238     & V351 Ori  & 05:44:18.8 & +00:08:40 & FIRE & 2016 Sept 20 & 1 $\times 4$ \\ 
HD 50083     & V742 Mon  & 06:51:45.8 & +05:05:04 & GNIRS & 2012 Dec 13 & 90 \\
HD 52721     & GU~CMa    & 07:01:49.5 & --11:18:03 & FIRE & 2016 Sept 20 & 2 $\times 4$ \\ 
HD~53367     & MWC 166   & 07:04:25.5 & --10:27:16 & FIRE & 2011 Mar 10 & 1 $\times 4$ \\ 
HD 68695     &           & 08:11:44.3 & --44:05:08 & X-shooter & 2009 Dec 21 & 15 $\times 4$ \\
HD 72106     &           & 08:29:35.0 & --38:36:19 & X-shooter & 2009 Dec 19 & 10 $\times 4$ \\
HD 76534     &           & 08:55:08.8 & --43:27:57 & X-shooter & 2010 Jan 30 & 15 $\times 4$ \\ 
HD 98922     &           & 11:22:31.7 & --53:22:11 & FIRE & 2011 Mar 10 & 10 $\times 4$ \\ 
HD 101412    &           & 11:39:44.3 & --60:10:25 & X-shooter & 2010 Mar 30 & 10 $\times$ 4 \\ 
HD 114981    & V958 Cen  & 13:14:40.7 & --38:39:06 & GNIRS & 2013 Jan 16 & 180 \\
HD 139614    &           & 15:40:46.3 & --42:29:51 & X-shooter & 2010 Mar 28 & 5 $\times 4$ \\ 
HD 141569    &           & 15:49:57.7 & --03:55:16 & FIRE & 2011 Mar 12 & 20 $\times 4$ \\  
HD 142666    &           & 15:56:40.0 & --22:01:40 & FIRE & 2011 Mar 12 & 10 $\times 4$ \\   
HD 144432    &           & 16:06:58.0 & --27:43:10 & FIRE & 2011 Mar 12 & 5 $\times 4$ \\  
HR 5999      &           & 16:08:34.3 & --39:06:18 & FIRE & 2011 Mar 10 & 1 $\times 4$ \\ 
HD 145718    & V718 Sco  & 16:13:11.6 & --22:29:07 & PHOENIX & 2013 Feb 27 & 600 \\
HD~150193    & MWC 863   & 16:40:17.9 & --23:53:45 & FIRE & 2011 Mar 10 & 1 $\times 4$ \\ 
HD 152404    & AK Sco    & 16:54:45.0 & --36:53:17 & X-shooter & 2009 Oct 05 & 5 $\times 4$ \\ 
HD 163296    &           & 17:56:21.3 & --21:57:22 & FIRE & 2011 Mar 10 & 10 $\times 4$ \\ 
HD 169142    &           & 18:24:29.8 & --29:46:49 & FIRE & 2016 Sept 20 & 2 $\times 4$ \\ 
HD 174571    & MWC 610   & 18:50:47.2 & +08:42:10 & FIRE & 2016 Sept 20 & 3 $\times 4$ \\ 
HD 176386    &           & 19:01:38.9 & --36:53:26 & FIRE & 2016 Sept 20 & 1 $\times 4$ \\ 
HD~179218    & MWC 614   & 19:11:11.3 & +15:47:16 & FIRE & 2011 Mar 12 & 1 $\times 4$ \\ 
HD~190073    & V1295 Aql & 20:03:02.5 & +05:44:17 & FIRE & 2011 Mar 12 & 3 $\times 4$ \\ 
HD 200775    & MWC 361   & 21:01:36.9 & +68:09:48 & PHOENIX & 2013 Feb 27 & 300 \\
HD~216629    & IL Cep    & 22:53:15.6 & +62:08:45 & PHOENIX & 2013 Nov 9 & 600 \\
HD 244314    & V1409 Ori & 05:30:19.0 & +11:20:20 & X-shooter & 2010 Jan 02 & 15 $\times 4$ \\
HD 244604    & V1410 Ori & 05:31:57.2 & +11:17:41 & PHOENIX & 2013 Feb 28 & 600 \\
HD 245185    & V1271 Ori & 05:35:09.6 & +10:01:52 & X-shooter & 2009 Dec 17 & 15 $\times 4$ \\
HD 249879    &           & 05:58:55.8 & +16:39:57 & FIRE & 2016 Sept 20 & 5 $\times 4$ \\ 
HD 250550    & V1307 Ori & 06:01:60.0 & +16:30:57 & PHOENIX & 2013 Feb 27 & 700 \\
HD~259431    & MWC 147   & 06:33:05.2 & +10:19:20 & FIRE & 2011 Mar 10 & 1 $\times 4$ \\ 
HD 275877    & XY Per    & 03:49:36.3 & +38:58:56 & PHOENIX & 2013 Nov 8 & 600 \\
HD 278937    & IP Per    &  03:40:47.0 & +32:31:54 & PHOENIX & 2013 Feb 28 & 800 \\
HD 287823    &           & 05:24:08.0 & +02:27:47 & PHOENIX & 2013 Mar 02 & 600 \\
HD 287841    & V346 Ori  & 05:24:42.8 & +01:43:48 & PHOENIX & 2013 Nov 08 & 600 \\
HD 290409    &           & 05:27:05.5 & +00:25:08 & X-shooter & 2010 Jan 02 & 15 $\times 4$ \\
HD 290500    &           & 05:29:48.0 & --00:23:43 & X-shooter & 2009 Dec 17 & 75 $\times 4$ \\
HD 290770    &           & 05:37:02.4 & --01:37:21 & X-shooter & 2009 Dec 26 & 7 $\times 4$ \\
HD 293782    & UX Ori    & 05:04:29.9 & --03:47:14 & FIRE & 2016 Sept 20 & 2 $\times 4$ \\ 
             & MWC 1080  & 23:17:25.6 & +60:50:43 & PHOENIX & 2013 Nov 10 & 600 \\
             & VV Ser    & 18:28:47.9 & +00:08:39 & FIRE & 2016 Sept 20 & 3 $\times 4$ \\ 
             & LkHa 215  & 06:32:41.8 & +10:09:34 & PHOENIX & 2013 Mar 1 & 800 \\
\hline
\end{longtable*}
\end{center}
\end{footnotesize}


\subsection{Other spectra}\label{ss:other_spec}
In order to obtain He~{\sc i}~10830~\AA\ profiles for as many sources reported by \citet{ale13} as possible, we also include profiles from \citet{cau14} and \citet{fai15}. 
We list the observational details for these 32 spectra in Table~\ref{t:obs}.

Near-IR spectra from \citet{fai15} were obtained in 2009-2010 with the X-Shooter spectrograph on the VLT \citep{ver11}. 
We obtained Phase 3 pipeline-reduced spectra from the ESO data archive. 
Spectral resolution for the 0.4\arcsec-0.5\arcsec\ slit widths used by \citet{fai15} is $R \sim 10,500$. 
A more complete description of the data is given by \citet{fai15}.

We also include He~{\sc i}~10830~\AA\ profiles from \citet{cau14}.
These targets are primarily in the northern hemisphere and complement the sample of sources in the southern hemisphere obtained with FIRE and X-Shooter. 
Specific observation parameters for these data are described by \citet{cau14}. 
Briefly, data were obtained with GNIRS \citep{eli06} on Gemini ($R \sim 18,000$) and PHOENIX \citep{hin98} on the Mayall 4~m and KPNO 2.1~m ($R \sim 50,000$). 
Gratings and order-blocking filters were used to isolate emission near the He~{\sc i}~10830~\AA\ line. 
Data were reduced using custom \textsc{IDL} routines.

\begin{table*}[ht!]
\caption{Herbig Ae/Be stars with a detected B-field}
\vspace{5pt}
\centering
\footnotesize
\vspace{3pt}
\begin{tabular}{lllllll}
\hline\hline
\vspace{5pt}
HD number & Target & SpT & $log(L_{bol})$ & LSD$^*$ [G] & Regression$^*$ [G] & line profile \\ 
\hline 
BD-06 1253$^a$ & V380 Ori  & B9    & 1.99 & $-165 \pm 190$ & 460 $\pm$ 70 & PC \\
HD~36982$^b$   & LP~Ori    & B1.5  & 3.22 & 220$\pm$50 & & O \\
HD~190073$^c$  & V1295 Aql & A1    & 1.92 & 111$\pm$13 & 91$\pm$81$^{h}$ & PC \\
HD~101412$^d$  &           & B9/A0 & 1.36 & 785$\pm$55$^{g}$ & 465$\pm$27$^{f}$ & IPC \\ 
HD 35929$^{e,\dagger}$   &           & F1    & 2.12 & $74 \pm 19$ & & O \\
\hline
\multicolumn{7}{l}{$^*$see discussion of reduction methods in Section~\ref{ss:obs_Bfields}} \\ 
\multicolumn{7}{l}{$^{\dagger}$ HD~35929 is a marginal detection and a $\delta$-Scuti pulsating variable \citep{mar00} } \\
\multicolumn{7}{l}{whose polarization signature may not be magnetic in origin. } \\ 
\multicolumn{7}{l}{B-field discovery papers: $^a$ \citet{wad05}, $^b$ \citet{pet08}, $^c$ \citet{cat07}, } \\ 
\multicolumn{7}{l}{$^d$ \citet{wad07}, $^e$ \citet{ale13}, $^{f}$ \citet{hub11}, $^{g}$ \citet{wad16}, } \\
\multicolumn{7}{l}{$^{h}$ \citet{hub13}} \\
\end{tabular} 
\label{t:detected_bfield}
\end{table*}

\subsection{B field measurements and source properties}\label{ss:obs_Bfields}
We list the stellar parameters -- M$_{\star}$, L$_{\star}$, R$_{\star}$, $v sin(i)$, $v_{rad}$ -- for each source in Table~\ref{t:demographics_table} using data from \citet{ale13} and \citet{fai15}.
Longitudinal field strengths for the sources with a detection are listed in Table~\ref{t:detected_bfield}. 
Deriving a field topology from the longitudinal field strength requires time-series measurements and models of the surface magnetic field \citep[e.g.,][]{don07}. 
We do not include sources where the magnetic field is likely associated with a low-mass companion, i.e., HD~72106 and HD~200775 \citep[][respectively]{fol08,ale08}. 
A few of these objects also have magnetic field detections reported by \citet{hub09,hub11}; we include these in Table~\ref{t:detected_bfield} where available. 
Both groups use spectropolarimetric data to determine the longitudinal field strength. 
  Alecian et al.\ measure magnetic fields using the Least-Square Deconvolution (LSD) technique with higher-resolution ($R \simeq 65000$) data from ESPaDOnS (on CFHT) and/or Narval (on T\'{e}lescope Bernard Lyot).
Hubrig et al.\ use the regression method of \citet{bag02} on lower-resolution ($R \simeq 2000$) FORS (on VLT) data. 
The two groups often come to different conclusions about the magnetic field strength \citep[e.g.,][]{hub11}. 
\citet{ale13} apply a more stringent detection criterion by computing a false alarm probability that the observed Stokes V profile could be produced in the absence of a magnetic field. 
In contrast, Hubrig et al.\ assume that photon counting statistics dominate their uncertainties. 
However, \citet{bag12} demonstrated that photon noise is not the only source of uncertainty even in high signal-to-noise data. 
Re-reducing the FORS data with a more complete treatment of the uncertainties, \citet{bag12} do not confirm many of the field detections claimed in the literature. 
For our analysis, we prefer the more conservative criteria employed by \citet{ale13}, although we also include field strengths for those stars measured by Hubrig et al.\ in Table~\ref{t:detected_bfield}.

Spectral types, mass accretion rates, $\dot{M}_{acc}$, and disk inclinations, $cos(i)$, taken from the literature are listed 
in Table~\ref{t:demographics_table}. 
Spectral types are from Table~2 in \citet{cau14}
or Simbad. 
Mass accretion rates are taken from \citet{fai15} where available, or from the literature via Table~2 in \citet{cau14}. 
Most disk inclination angle estimates are taken from \citet{laz17} who use PIONIER/VLTI to obtain H-band ($1.4-1.8$~\micron) long-baseline interferometric observations of HAeBe disks. 
Near-IR visibilities are fit with either an ellipsoidal or ring-like profile.
For the values listed in Table~\ref{t:demographics_table}, we report inclinations derived from the ring-like (labeled ``rl'' in Table~\ref{t:demographics_table}) model where available, but include those derived from the ellipsoidal (``el'' in Table~\ref{t:demographics_table}) model for sources where no value for the ring model is reported. 
Following the inclination angle conventions in that paper, a source viewed edge-on will have $cos(i) = 0$ while a source viewed pole-on will have $cos(i) = 1$.

\section{Line Profiles}\label{s:results}
\begin{figure*}
\centering
\includegraphics[trim=10mm 5mm 10mm 5mm,angle=0,scale=0.875]{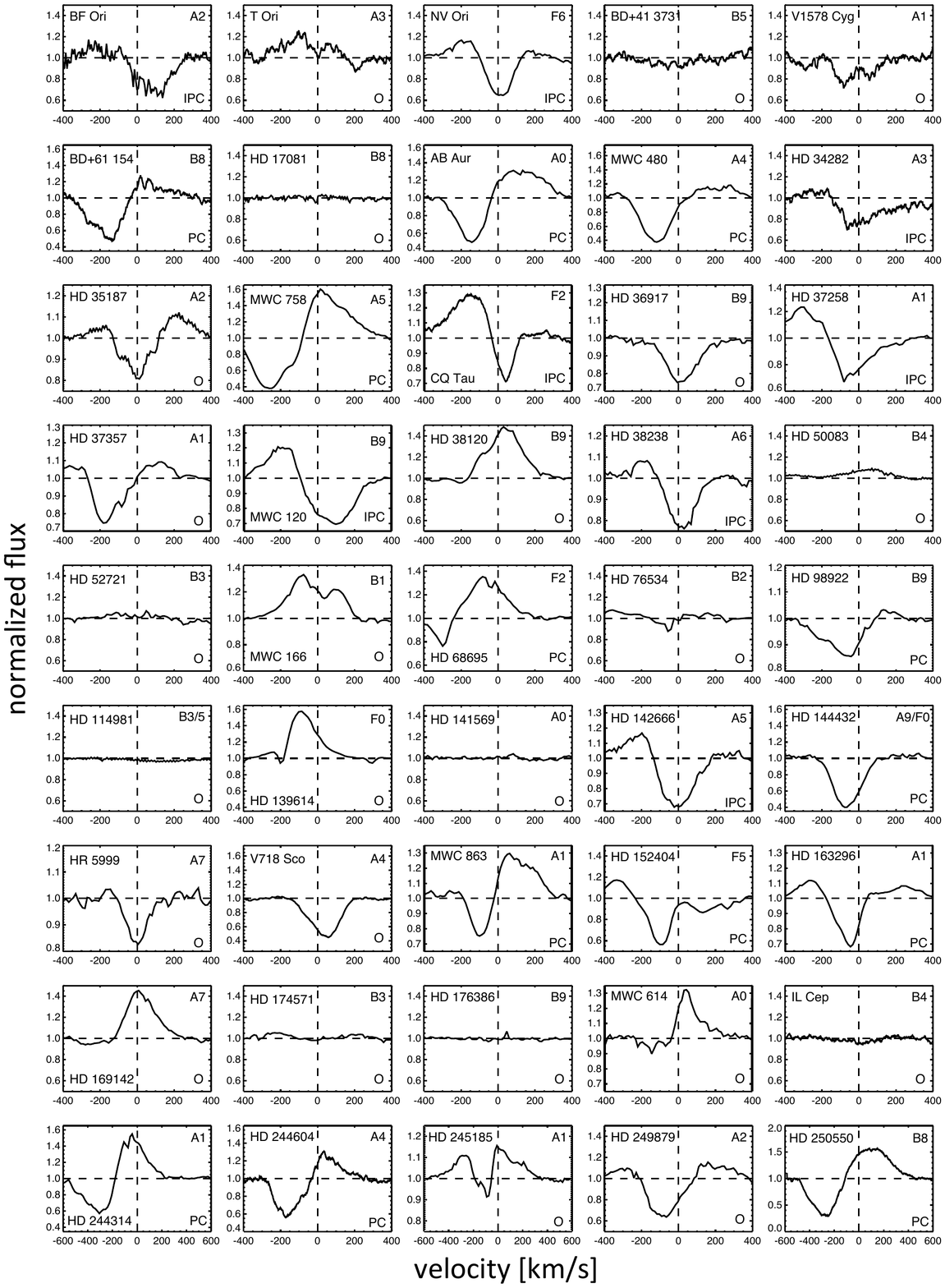}
\end{figure*}
\begin{figure*}[t!]
\centering
\includegraphics[trim=10mm 0mm 10mm 15mm,angle=0,scale=0.875]{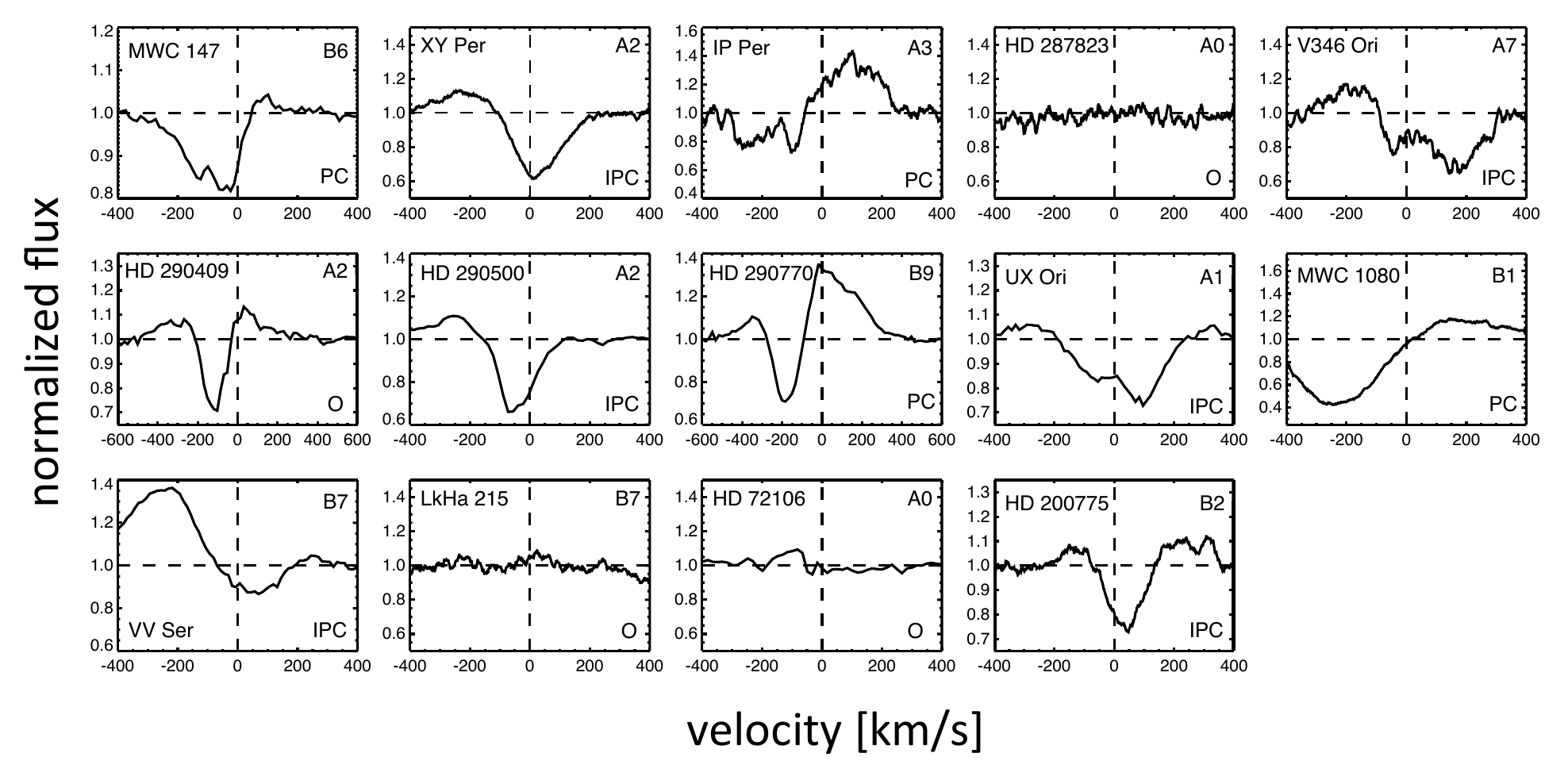} 
\caption{He~{\sc i}~10830~\AA\ line profiles of sources targets for a magnetic field measurement by \citet{ale13}. 
}\label{fig:HeI_profiles} 
\end{figure*}
\begin{figure*}
\centering
\includegraphics[trim=0mm 0mm 0mm 50mm,angle=0,scale=0.975]{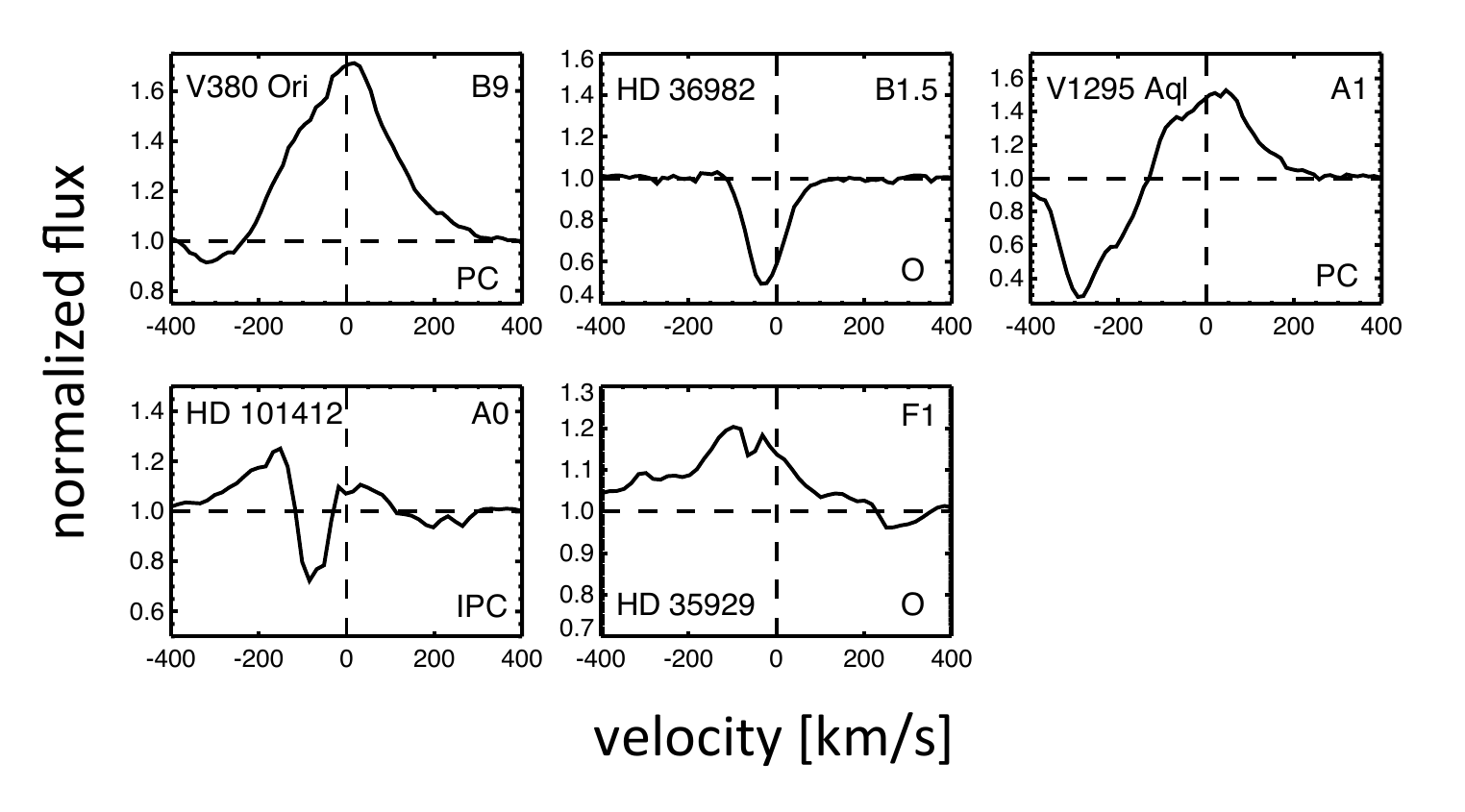} 
\caption{Same as Figure~\ref{fig:HeI_profiles}, but for HAeBes with a magnetic field detection (see Table~\ref{t:detected_bfield}). 
}\label{fig:HeI_profiles_Bdetected} 
\end{figure*}
%


The sensitivity of various spectral lines to the kinematics of gas near young stars depends on the optical depth and thermalization of the line \citep[see, e.g.,][]{cal92,hartmann94,muz01}. 
HAeBes are hotter and brighter than T~Tauri stars and the Balmer lines tend to be more optically thick, making it difficult to trace the kinematics of the gas from the line profile \citep{muz04}.
This is especially true if targeting redshifted absorption, as we do here, as an unambiguous indicator of infalling gas.

Instead of the Balmer lines, 
we examine the profile of He~{\sc i}~10830~\AA\ to look for signatures of accretion (i.e.\ redshifted absorption) in HAeBes targeted for a measurement of the magnetic field by \citet{ale13}. 
He~{\sc i}~10830~\AA\ has not yet been calibrated as a measure of the accretion rate. 
However, the metastable lower level makes the transition particularly prone to absorption, and therefore a sensitive tracer of the kinematics of gas near the star. 
The line is primarily populated through recombinations following excitation by high-energy photons near the star \citep[e.g.,][]{dup05,kur11}. 
The lower level lies 20~eV above ground, restricting the region of its formation -- and therefore the gas kinematics it traces --- closer to the star.

Combining our new FIRE spectra with those from the literature, we have He~{\sc i}~10830~\AA\ profiles for 
63/70 (90\%) of the HAeBes targeted for a magnetic field measurement by \citet{ale13}. 
We also include a previous confirmation from the literature (HD~101412) for a total 64 HAeBes with spectra and 5/64 (8\%) with a detected magnetic field. 
We present the He~{\sc i}~10830~\AA\ profiles for all 64 sources in Figures~\ref{fig:HeI_profiles}-\ref{fig:HeI_profiles_Bdetected}. 
Profile shape classifications for each source are listed in Table~\ref{t:demographics_table}. 
Since we are primarily interested in tracers of accretion, we follow a simplified version of the line classification scheme used by \citet{cau14}, sorting profiles into one of three categories: 
(1) blueshifted absorption (P~Cygni profiles -- PC; 19 or 30\%), 
(2) redshifted absorption (inverse P~Cygni profiles -- IPC; 15 or 23\%), or 
(3) other profile type (O; 30 or 47\%).

Altogether, 53\% of the line profiles show redshifted or blueshifted absorption, either directly or indirectly (see Section~\ref{s:discussion}) indicative of disk accretion. 
A similar fraction of T~Tauri stars display PC or IPC profiles, with half (18/38; 47\%) of the sources presented by \citet{edw06} showing one of the two profiles.
However, considering all profiles that show redshifted absorption
(even if blueshifted absorption is also present), \citet{edw06} find that 
half the sample shows redshifted absorption in He~{\sc i}~10830~\AA.
A more complete analysis of redshifted absorption in HAeBes requires detailed modeling of the excitation and radiative transfer driving He~{\sc i}~10830~\AA\ line formation \citep[see, e.g.,][]{fis08,kur11}.

Among the 5 HAeBes with a reported magnetic field detection (see Table~\ref{t:detected_bfield}), 
roughly half (2/5) show blueshifted absorption (PC), one has redshifted absorption (IPC), and the remaining two have other profiles shapes.
  Differences in the line profiles between the magnetic HAeBes and the non-magnetic HAeBes are difficult to quantify, given the small number of HAeBes with a detected magnetic field.

\begin{figure*}
\centering
$\begin{array}{cc}
\includegraphics[trim=10mm 10mm 10mm 10mm,angle=90,scale=0.325]{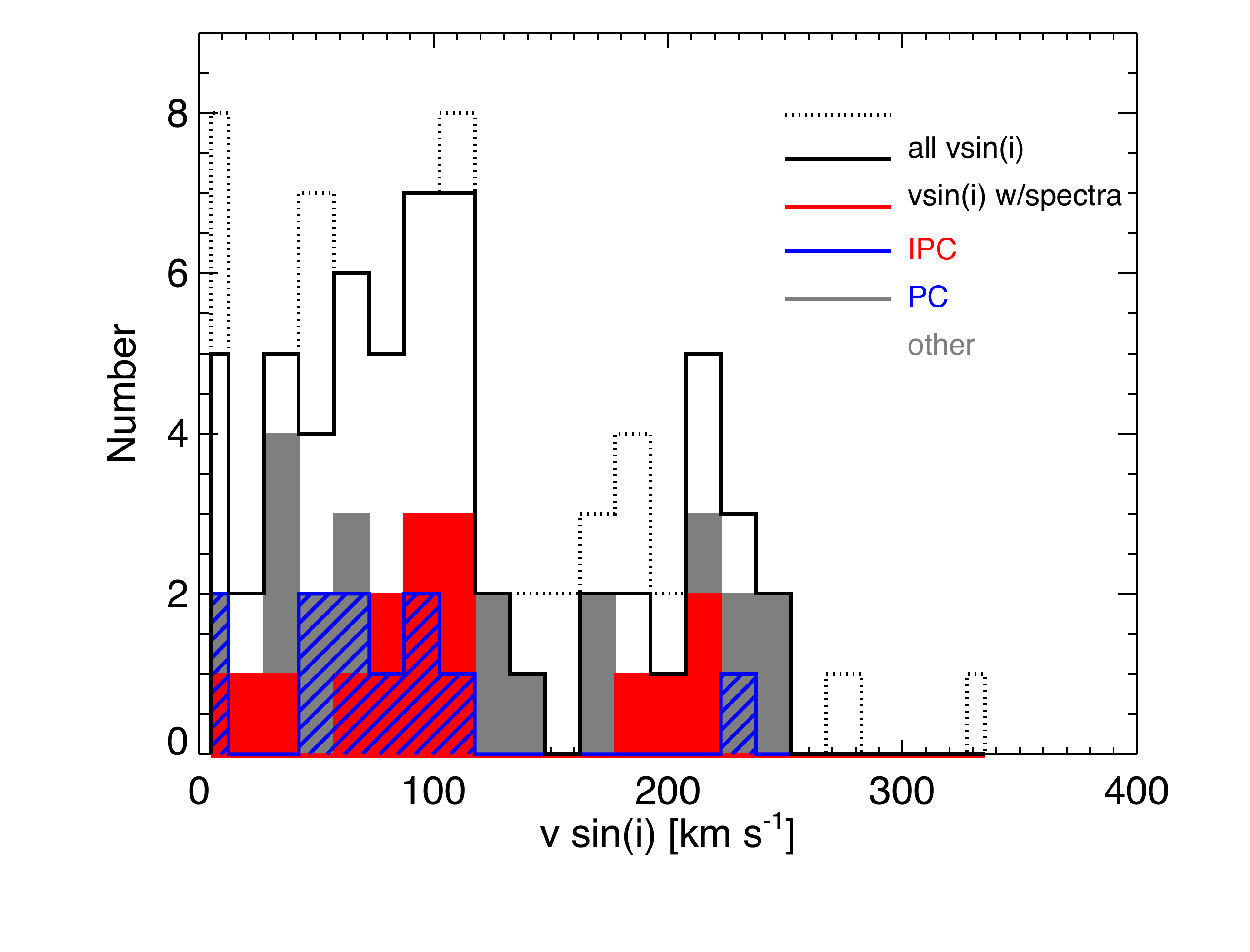} & 
\includegraphics[trim=10mm 10mm 10mm 10mm,angle=90,scale=0.325]{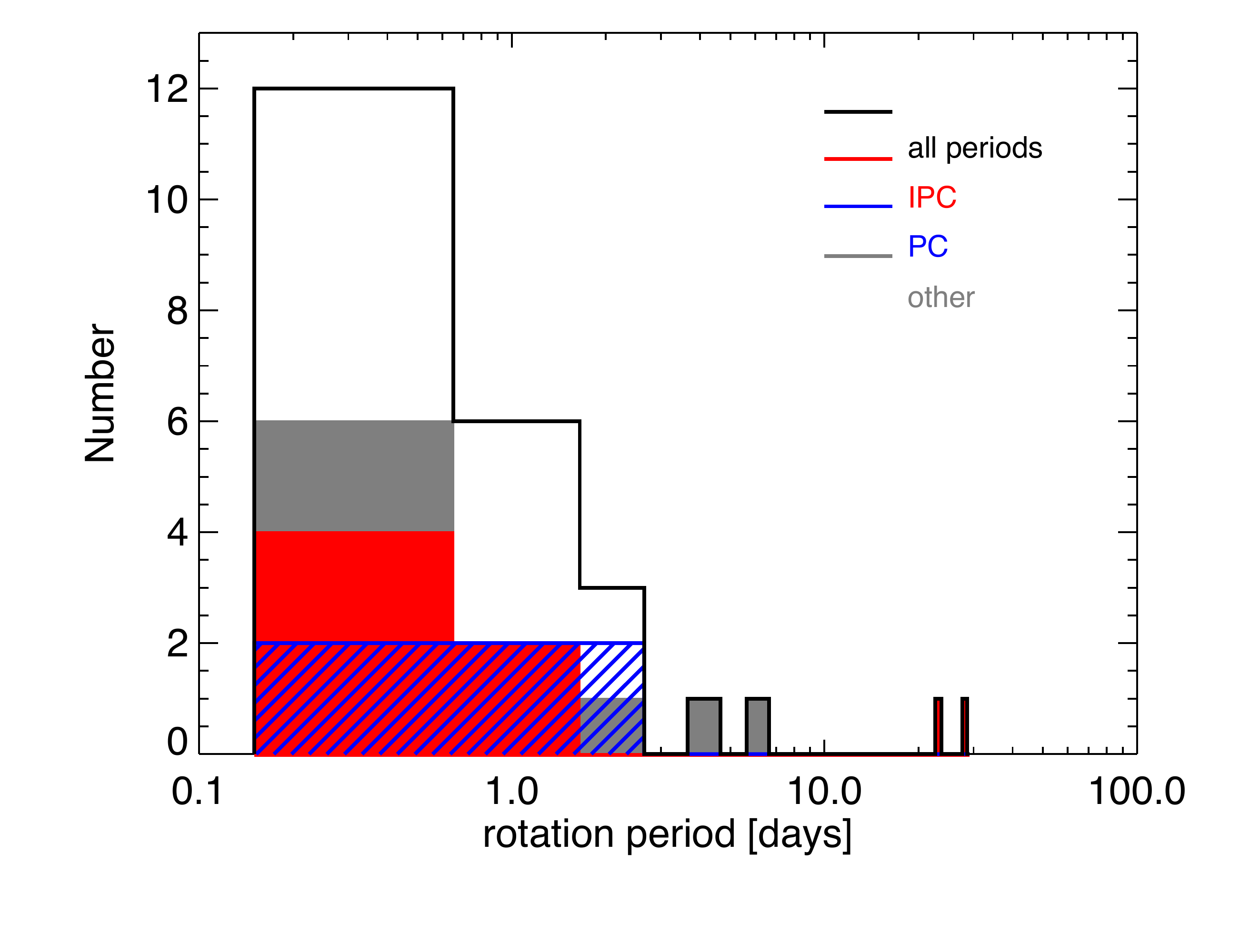} \\
\includegraphics[trim=10mm 10mm 10mm 10mm,angle=90,scale=0.325]{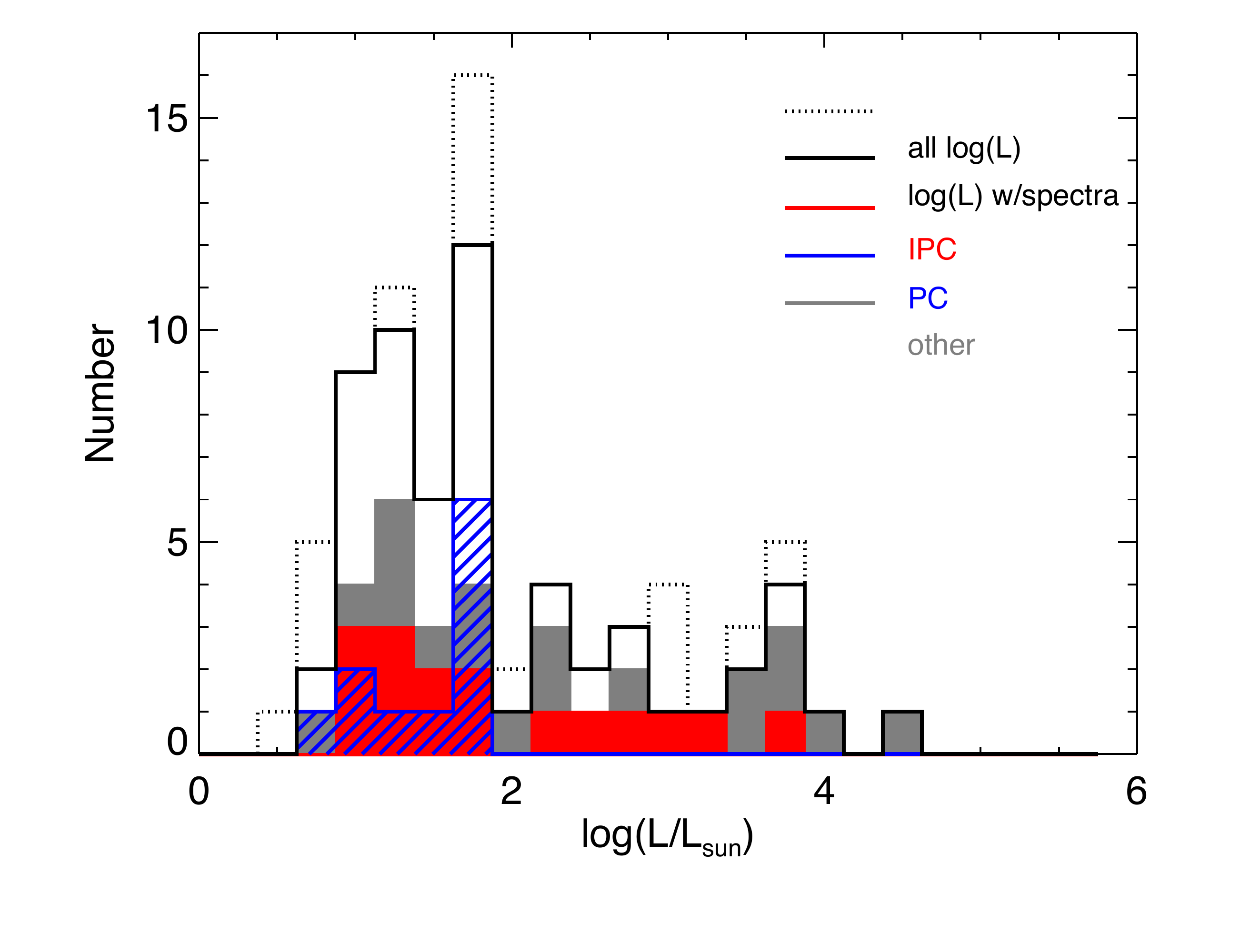} &
\includegraphics[trim=10mm 10mm 10mm 10mm,angle=90,scale=0.325]{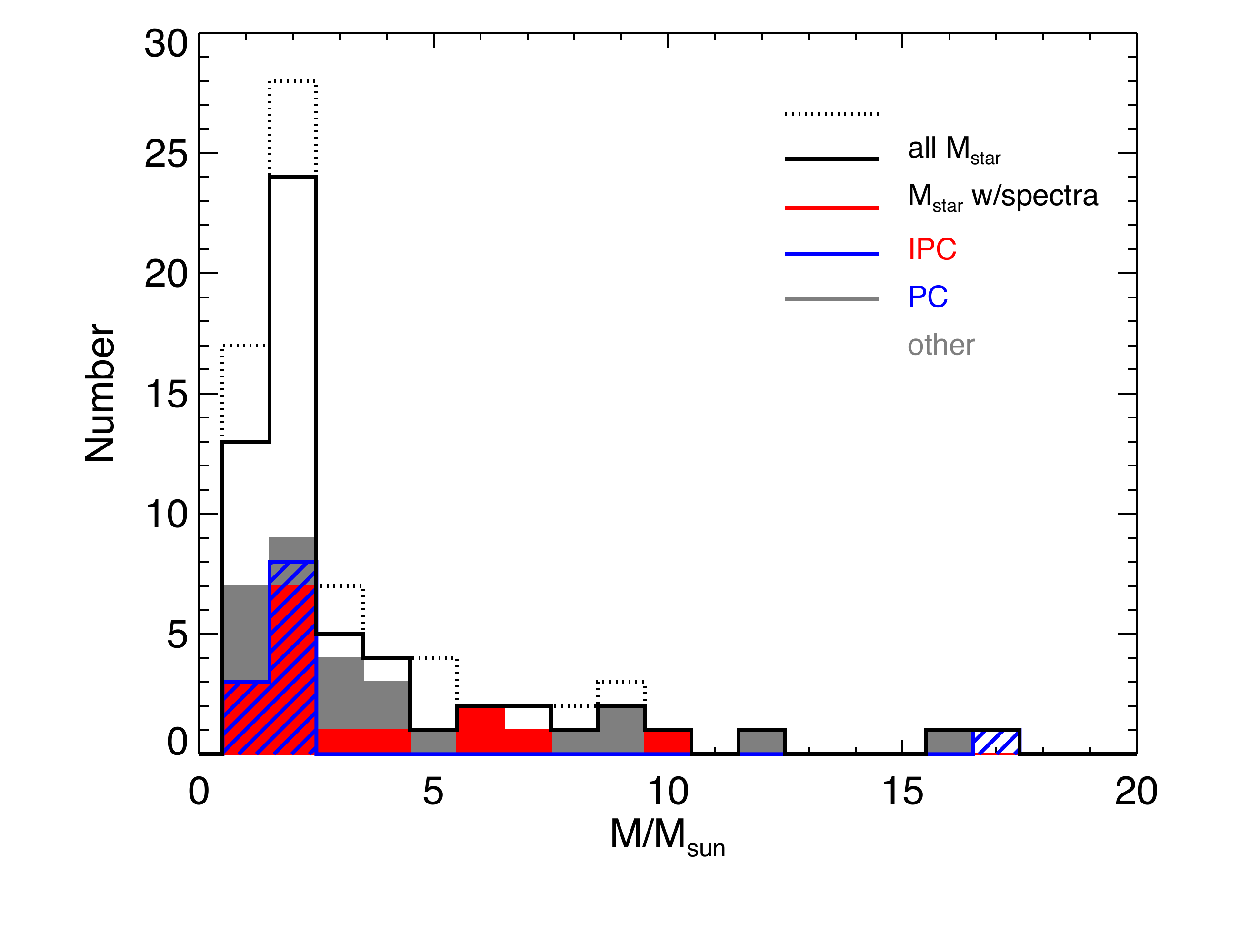} \\
\includegraphics[trim=10mm 10mm 10mm 10mm,angle=90,scale=0.325]{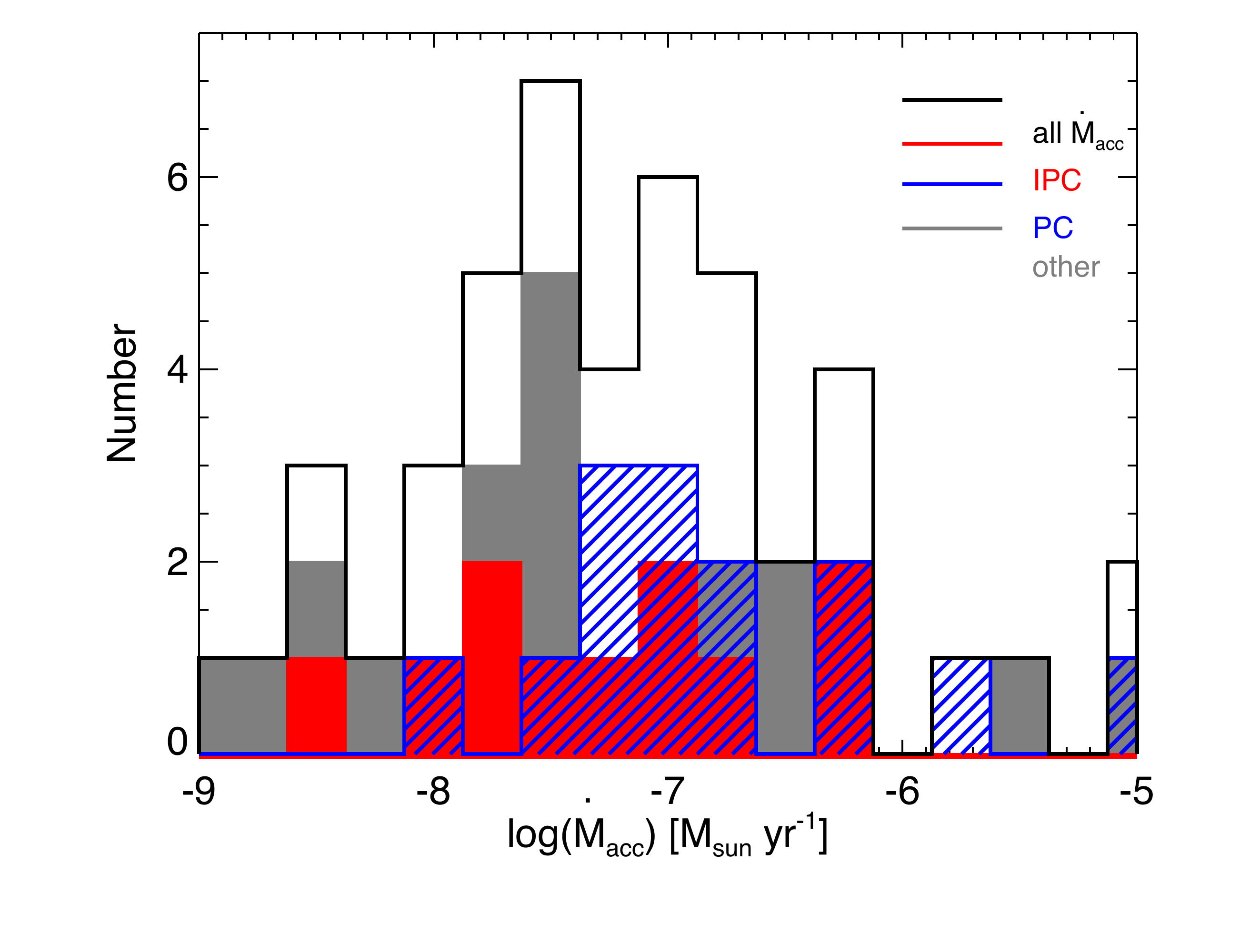} &
\includegraphics[trim=10mm 10mm 10mm 10mm,angle=90,scale=0.325]{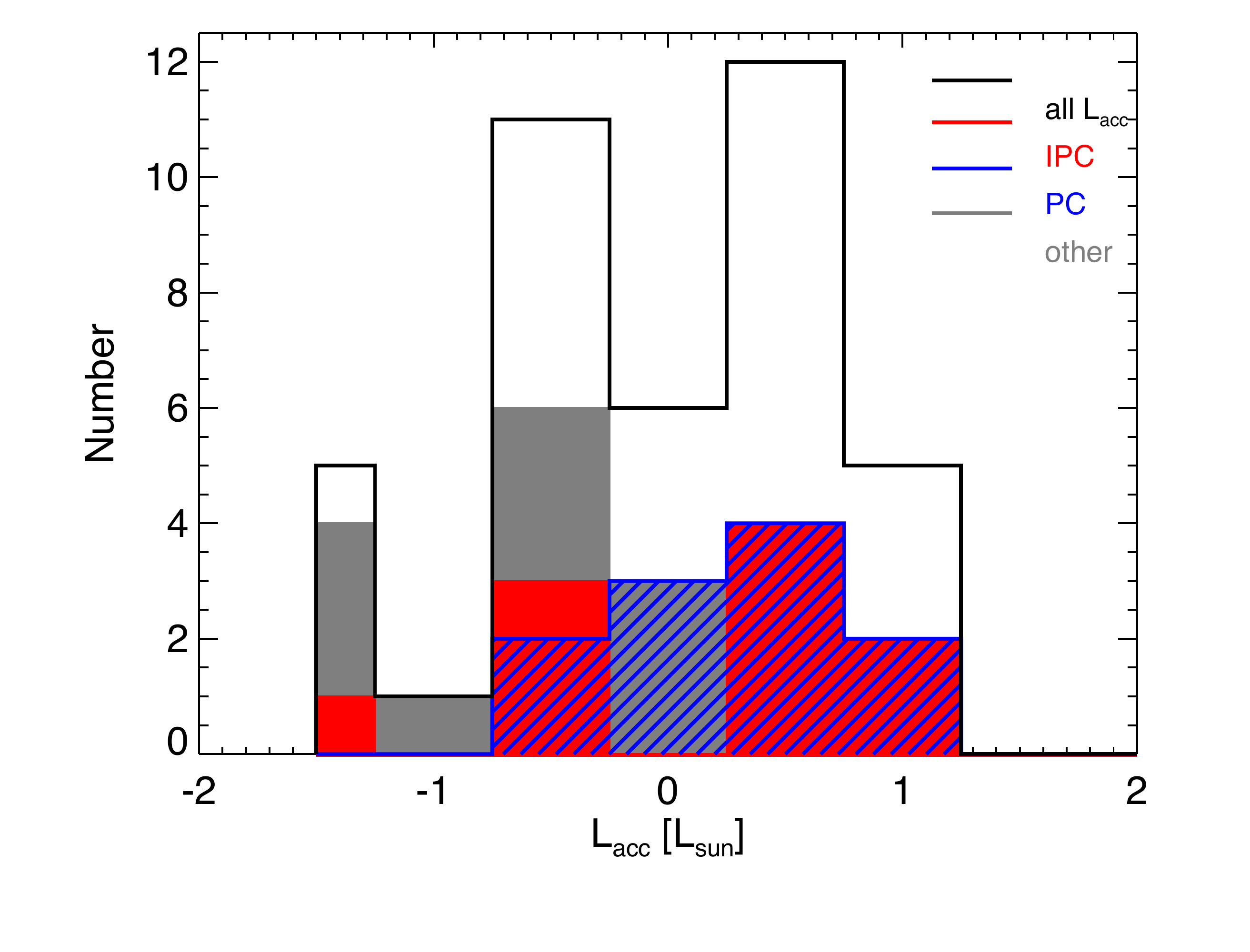} \\
\end{array}$
\caption{Histograms showing He~{\sc i}~10830~\AA\ line profiles observed as a function of stellar parameters (see Table~\ref{t:demographics_table}). 
The dotted line shows all sources observed by \citet{ale13}, the solid line indicates the subset with He~{\sc i}~10830~\AA\ data. 
The red histogram shows those sources with redshifted absorption; blue dashed histogram shows the sources with blueshifted absorption; gray histogram shows the distribution of all other line profiles classifications. 
}\label{fig:profile_hist} 
\end{figure*}

Magnetic fields are notoriously difficult to measure \citep{sho02}. 
We therefore conduct a parameter study to explore whether line profile morphology correlates with any of the stellar parameters, indirectly indicating the influence of the magnetic field. 
Figure~\ref{fig:profile_hist} shows the distribution of redshifted absorption, blueshifted absorption, and other line profiles as a function of the stellar parameters listed in Table~\ref{t:demographics_table}.

Magnetic braking has been invoked to explain the slow rotation of some stars. 
Indeed, the magnetic HAeBes appear to have much slower rotation rates than the non-magnetic sample \citep{ale13b}. 
At the same time, magnetic fields are more difficult to measure in stars with faster rotation rates, as typically observed in HAeBes \citep{sho02}. 
Faster rotation rates may obscure Zeeman broadening due to more modest magnetic fields, leading to the preferential detection in sources where fields are anomalously large or the rotation rates are particularly slow. 
Nevertheless, 
if slower rotators are more likely to be associated with redshifted absorption, this may be interpreted as indirect evidence for a magnetic field. 
We show a histogram of different line profiles versus $v sin(i)$ in Figure~\ref{fig:profile_hist}. 
Neither the distribution of redshifted absorption nor blueshifted absorption are statistically different from the overall distribution of $v sin(i)$.


%


The true rotation rate is likely larger than the measured $v sin(i)$ since most sources will not be observed edge-on. 
Uncertainty in the viewing angle will redistribute fast and slow rotators in a plot of $v sin(i)$. 
Some of this uncertainty can be mitigated where inclination constraints exist. 
Disk inclinations for 25 of the targets considered here were recently derived from H-band interferometry by \citet{laz17}.
We list these in Table~\ref{t:demographics_table}, together with 
an inclination estimate for HD~101412 from the literature. 
We assume that the disk axis and the rotation axis of the star are aligned.
This allows us to compute the rotation period for stars with disk inclination estimate using the radii listed in Table~\ref{t:demographics_table}. 
Rotation periods for these 27 stars are listed in Table~\ref{t:demographics_table} and 
the resulting distribution of line profiles as a function of period is shown in Figure~\ref{fig:profile_hist}.

\begin{figure}
\centering
$\begin{array}{c}
\includegraphics[trim=10mm 0mm 0mm 10mm,angle=90,scale=0.35]{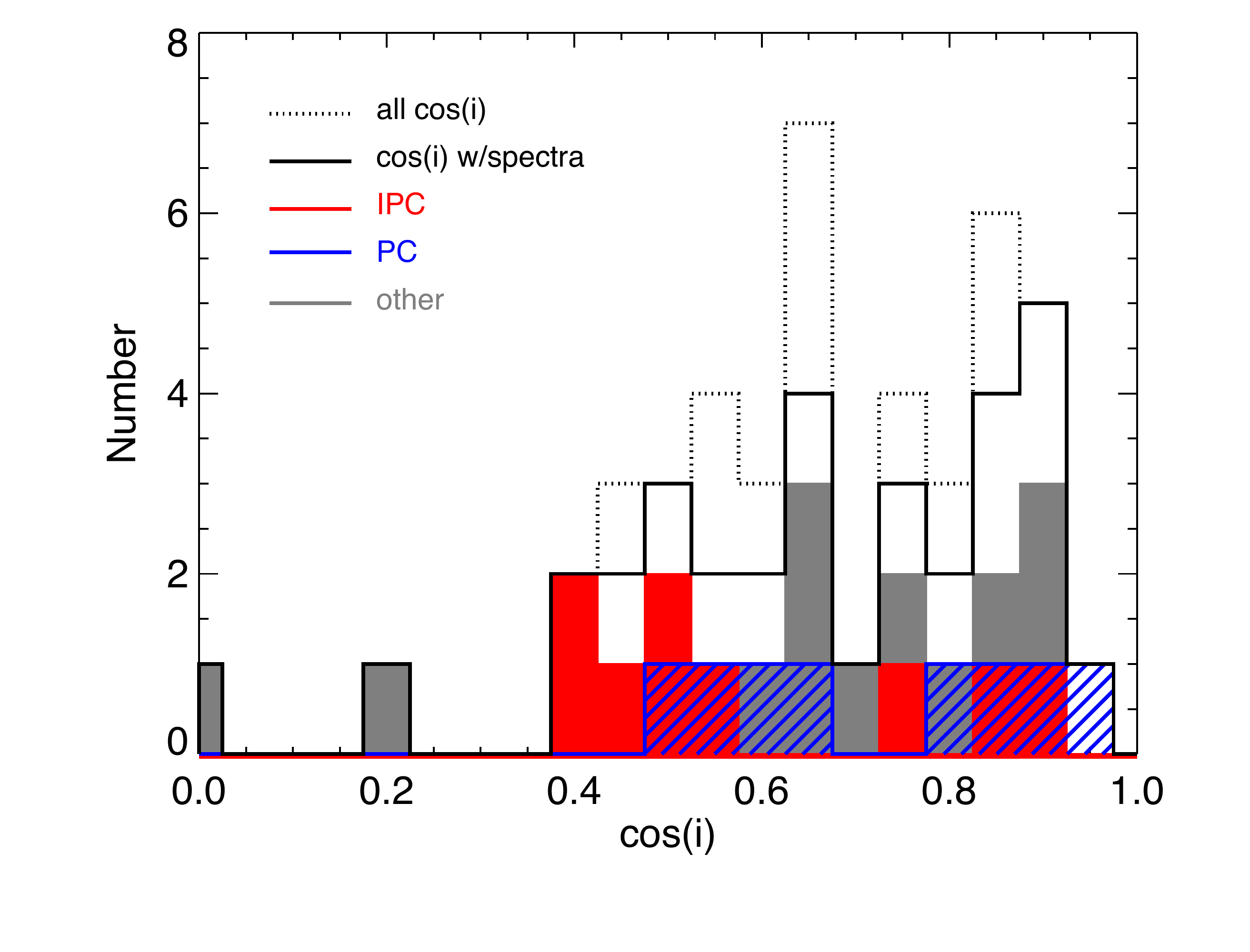} \\
\includegraphics[trim=10mm 0mm 0mm 10mm,angle=90,scale=0.35]{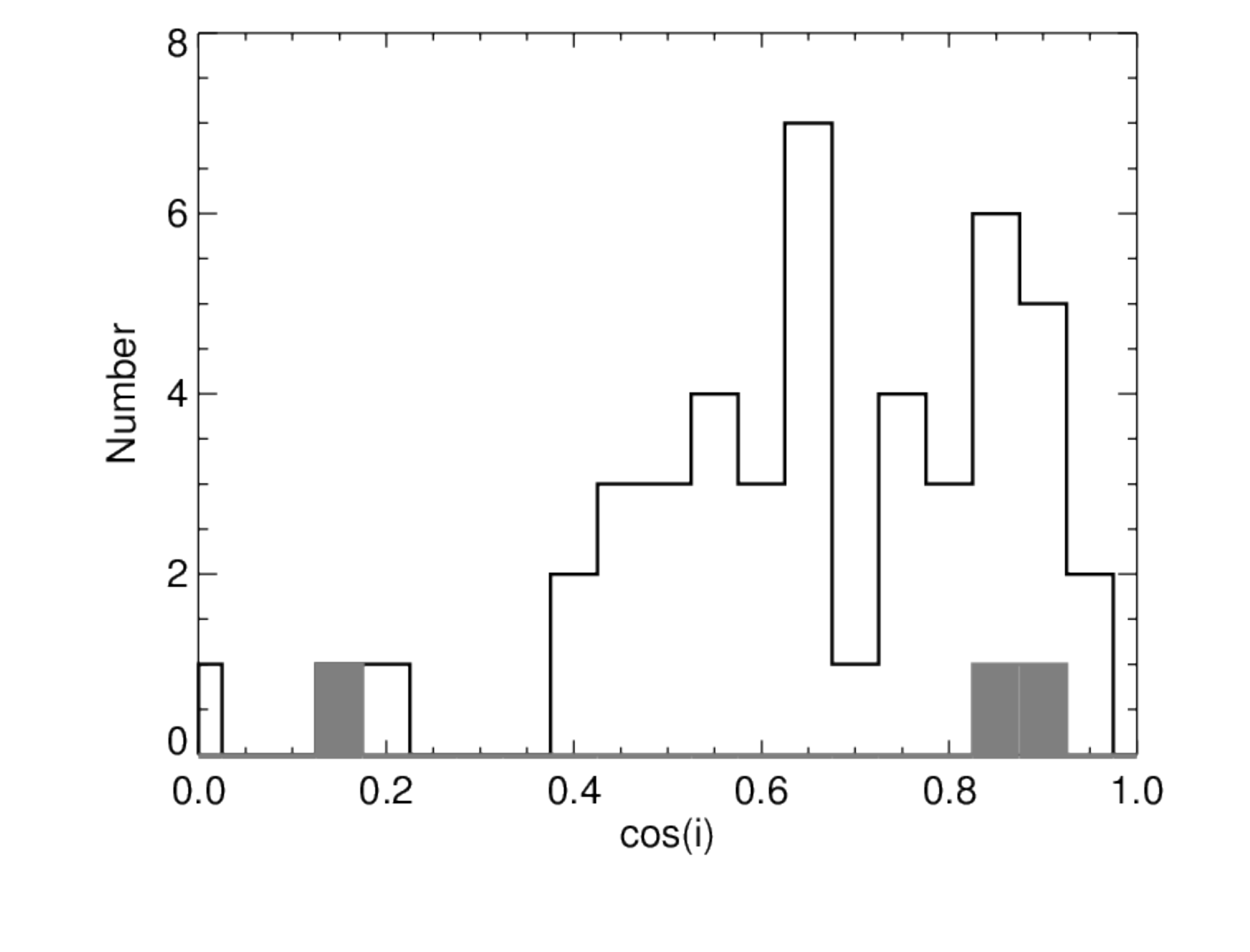} \\ 
\end{array}$
\caption{\textit{Top:} Histograms showing line profiles observed in sources with an estimated disk inclination angle. 
As in Figure~\ref{fig:profile_hist}, the dotted line shows all sources with an inclination estimate, the solid line indicates the subset with a spectrum, 
the red histogram shows those sources with redshifted absorption, the blue dashed histogram shows blueshifted absorption profiles, and the gray histogram shows all other line profiles classifications. 
\textit{Bottom:} Histogram showing the source inclination angles of magnetic HAeBes (gray) compared to the overall sample with estimated inclination angles.  
The source with $cos(i)<0.5$ is HD~101412 (see Section~\ref{s:discussion}). 
}\label{fig:incl_Bfield} 
\end{figure}

We also consider line profiles as a function of the source inclination. 
In the magnetospheric accretion paradigm, redshifted absorption will only be observable for certain viewing geometries, i.e.\ if the accretion column intersects the line of sight \citep[see, e.g., Figure~1 in][]{hartmann16}. 
For sources that are viewed nearly pole-on, the line of sight is most likely to intersect the outflowing gas, and thus the line may be more likely to show blueshifted absorption. 
We plot line profiles as a function of the inclination angle in Figure~\ref{fig:incl_Bfield}. 
Among the small number of sources with both an inclination estimate and red/blueshifted absorption, there is no clear trend with source inclination.

For all stellar parameters that we consider, neither the redshifted nor the blueshifted absorption profiles are distributed in a manner that is statistically distinguishable from the overall sample. 
Probabilities returned from a two-sided Kolmogorov-Smirnov test for each parameter support the null hypothesis (that the two distributions have the same parent population). 
With the caveat that sample sizes are small (especially for parameters like $cos(i)$), none of the line profiles differ from the overall distribution with likelihood $\geq 90$\%. 

Lastly, we compare the inclination angles of the sample as a whole with the inclination angles of the magnetic HAeBes. 
The resulting histogram is shown in Figure~\ref{fig:incl_Bfield}. 
Few of the magnetic HAeBes also have a measured inclination. However, those that do tend to be observed nearly pole-on ($cos(i)>0.8$). 
The only magnetic HAeBe observed with a more edge-on orientation is HD~101412, with an estimated $cos(i) \approx 0.17$ from \citet{fed08}. 
However, this result derives from longer wavelength (mid-IR) observations obtained with significantly fewer baselines than used by \citet{laz17}. 
Marginally resolved observations of colder dust may sample different structures in the disk (i.e.\ flaring) at larger radii, biasing the inferred inclination angle.

\section{Discussion and Conclusions}\label{s:discussion}

We present He~{\sc i}~10830~\AA\ profiles of 64 HAeBes targeted for a magnetic field measurement. 
More than half of the sources in our sample display either redshifted (15/64; 23\%) or blueshifted (19/64; 30\%) absorption, directly or indirectly indicating accretion. 
\citet{cau14} find a remarkably similar proportion of IPC (redshifted) and PC (blueshifted) profiles (20\% and 30\%, respectively) in a similarly sized sample of HAeBes. 
While there is some overlap between the \citet{cau14} sample and the one presented here (see Table~\ref{t:obs}), the two studies differ in their target selection criterion. 
\citet{cau14} required only that a previous survey identified an object as a HAeBe whereas we present HAeBes that \citet{ale13} targeted for a magnetic field measurement. 
By comparing the profiles of sources with and without a detected magnetic field, we examine the role of strong, ordered magnetic fields in accretion in HAeBes.

Of the 64 HAeBes presented here, 5 (8\%) have a detected magnetic field (see Table~\ref{t:detected_bfield}). 
Most (3/5; 60\%) of the magnetic HAeBes show redshifted (1/5; 20\%) or blueshifted (2/5; 40\%) absorption in He~{\sc i}~10830~\AA. 
Redshifted absorption can only be created by infall and is the most direct indication that these HAeBes (magnetic or not) are accreting.

Blueshifted absorption tracing outflowing gas may be interpreted as indirect evidence for accretion. 
Accretion energy powers outflows and indeed sources with an enhanced accretion rate (i.e.\ FU~Ori-like outbursts) also have an elevated mass-loss rate \citep[e.g.,][]{cro87}.
However, the existence of an outflow cannot be taken as evidence of \textit{magnetospheric} accretion. 
Jet rotation studies \citep[e.g.,][]{bac02,cof04,fer06} suggest that jets may launch from a range of disk radii that lie outside the influence of the stellar magnetic field \citep[although see][]{cof15}. 
\citet{cau14} argue that a lack of narrow, blueshifted absorption in He~{\sc i}~10830~\AA\ points to the absence of inner disk winds. 
Instead, the broad blueshifted absorption profiles are more consistent with stellar winds \citep[e.g.,][Aarnio et al., submitted]{cat07}.

The He~{\sc i}~10830~\AA\ line profiles of the 64 HAeBes presented here are not significantly different between sources with and without a detected magnetic field. 
Roughly the same fraction of non-magnetic HAeBes show blueshifted absorption (17/59; 29\%) or redshifted absorption (14/59; 24\%) as the few sources with a detected field (40\% or 20\%, respectively).
A more robust comparison is difficult given the small number of sources with a field detection.

Despite the similarity of their line profiles, 
\citet{ale13b} report markedly different rotation rates between the magnetic and non-magnetic HAeBes. 
Magnetic braking \citep[e.g.,][]{ste00} may lead to slower rotation rates even in sources with fields too weak and/or too disordered to be detected. 
Coupling between the star and disk may be more difficult in this case; nevertheless, if the slowest rotating non-magnetic stars all show redshifted absorption, this might hint at an underlying magnetospheric accretion process. 
To test this hypothesis, we compare the He~{\sc i}~10830~\AA\ profiles to the projected rotational velocity measured at the stellar surface, $v sin(i)$ \citep[from][see Figure~\ref{fig:profile_hist}]{ale13}. 
The distribution of redshifted and blueshifted profiles are statistically indistinguishable from the overall distribution of $v sin(i)$ for the full sample.

Projection effects may intermix faster and slower rotators in a plot of $v sin(i)$, obscuring any underlying trend.
Assuming that the disk and rotation axes are aligned, we correct for projection and compute the rotation period for 26/64 HAeBes.
Again, the distributions of redshifted and blueshifted profiles do not indicate any trend with rotation period (see Figure~\ref{fig:profile_hist}). 
Altogether, this argues against accretion onto HAeBes being mediated by a dipolar magnetic field.

None of the line profiles appear to correlate with direct or indirect indications of a magnetic field.
Among the comparisons between stellar parameters and orientation effects that we explore in this paper, only the relationship between a magnetic field detection and the disk inclination of the source hints at a correlation. 
Of the three magnetic HAeBes with an estimated disk inclination (see Table~\ref{t:detected_bfield}), all but one have $cos(i) > 0.8$.
The only source seen nearly edge-on is HD~101412, although we consider this estimate less reliable than those obtained by \citet{laz17} (see Section~\ref{s:results}).

If inclination affects field detectability, this suggests that magnetic sensitivity is viewing angle dependent. 
Zeeman splitting of magnetic-sensitive lines will be harder to separate from Doppler broadening in stars with faster rotation rates. 
Stars seen nearly pole-on will have slower projected surface rotation velocities, and thus Doppler and Zeeman broadening may be more readily disentangled. 
Weak fields may be more difficult to detect among the faster rotating HAeBes.


\citet{gre12} have argued that the magnetic field evolves during the pre-main-sequence evolution of stars that develop a radiative envelope. 
All of the HAeBes in the \citet{ale13} sample lie in a portion of the H-R diagram where stars should have substantial radiative envelopes \citep[see Figure~4 in][]{gre12}. 
Whereas convective interiors can support the strong dipolar fields observed on some T~Tauris \citep[e.g.,][]{joh99a,joh07,joh13}, once the radiative envelope has developed, the dipole component gets weaker and the complexity of the field increases.

Even strongly magnetic low-mass stars are not well-described by a pure dipole \citep[e.g.,][]{joh99a}. 
Nevertheless, the strength of non-dipolar components decreases rapidly with distance from the stars \citep{val04}. 
While the standard magnetospheric accretion paradigm for low-mass stars assumes strong dipolar fields, several authors have argued that magnetic-mediated accretion can proceed even with realistically complex fields \citep[e.g.,][]{gre06,gre08,moh08,ada12,joh14}. 
Weak, complex fields may exist on HAeBes, below the detection limits of \citet{ale13}. 
The absence of strong dipolar fields does not preclude the possibility that accretion may proceed along the higher-order field lines that direct material closer to the stellar equator.

Comparing the He~{\sc i}~10830~\AA\ emission of HAeBes with and without a detected magnetic field suggests that line profile morphologies are insensitive to the magnetic field. 
However, few sources in our sample have a detected magnetic field; the majority have upper limits on the field strength. 
Both \citet{muz04} and \citet{cau14} argue that magnetically-mediated accretion may be possible through a more compact accretion geometry where higher-order field components create a smaller magnetosphere. 
Additional work is needed to model the excitation and radiative transfer of the He~{\sc i}~10830~\AA\ line.
Detailed theoretical models are essential to determine if the magnetospheric accretion paradigm can be modified to allow accretion through weaker, more topologically complex fields.

\acknowledgments
MR would like to thank Lee Hartmann and Chris Miller for useful discussions. 
JDM and AA were supported by NSF AST. 1311698. 
SK acknowledges support from an ERC Starting Grant (Grant Agreement No. 639889) and a STFC Rutherford fellowship (ST/J004030/1). 
This research has made use of the SIMBAD database, operated at CDS, Strasbourg, France. 
This is research has made use of the services of the ESO Science Archive Facility.
Based on observations collected at the European Organisation for Astronomical Research in the Southern Hemisphere under ESO programme(s) 088.C-0218(A).
Based on data obtained from the ESO Science Archive Facility under request numbers 231792, 231794, 231804, 231805, 231808, 231809, 231810, 255461, 231795, 231796, 231797, 231799, and 231802. 

%

\vspace{5mm}
\facilities{Magellan(FIRE), VLT(X-Shooter, Gemini(GNIRS), KPNO(PHOENIX)}



\bibliography{herbig_bibliography}


\begin{longrotatetable}
\begin{footnotesize}
\begin{center}
\begin{longtable*}{llllllllllll}
\caption{Herbig Ae/Be stars targeted for magnetic field measurement}\label{t:demographics_table} \\ 
\hline\hline
HD/BD & other & spectral & log(L$_{\star}$) & M$_{\star}$ & R$_{\star}$ & log($\dot{M}$) & 
$v sin(i)$ & cos(i) & v$_{eq}$ & $v_{rad}$ & He~{\sc i} \\ 
number & name & type & [L$_{\odot}$] & [M$_{\odot}$] & [R$_{\odot}$] & [M$_{\odot}$~yr$^{-1}$] & (km~s$^{-1}$) & 
       & (km~s$^{-1}$) & (km~s$^{-1}$) & Profile \\ 
\endfirsthead 
\hline 
BD-06~1259   & BF~Ori       & A2 & 1.75$^{+0.7}_{-0.7}$  & 2.58$^{+0.14}_{-0.14}$ & 3.26$^{+0.31}_{-0.31}$  & $-6.65^{+0.17}_{-0.25}$$^a$ & $39 \pm 9$ & ... &... & $22 \pm 6$ &  IPC \\  
BD-05~1253   & V380~Ori     & B9 & 1.99$^{+0.22}_{-0.22}$ & 2.87$^{+0.52}_{-0.32}$ & 3.00$^{+1.1}_{-0.8}$   & $-5.34^{+0.10}_{-0.15}$$^a$ & $6.7 \pm 1.1$ & ... &... & [27.3,28.2] &  PC \\ 
BD-05~1329   & T~Ori        & A3 & 1.97$^{+0.07}_{-0.07}$ & 3.13$^{+0.19}_{-0.19}$ & 4.47$^{+0.46}_{-0.46}$  & $-6.54^a$ & $147 \pm 9$ & ... &... & $29 \pm 8$ &  O \\ 
BD-05~1324   & NV~Ori       & F6$^*$ & 1.32$^{+0.07}_{-0.07}$ & 2.28$^{+0.18}_{-0.16}$ & 3.77$^{+0.41}_{-0.41}$  & $<-8.30^b$ & $74 \pm 7$ & ... &... & $30 \pm 5$ &  IPC \\ 
BD$+$41~3731 &              & B5 & 3.03$^{+0.36}_{-0.20}$ & 5.50$^{+1.37}_{-0.38}$ & 3.8$^{+0.8}_{-0.8}$   & ... & $345 \pm 27$ & ... &... & $-14 \pm 22$ &  O  \\ 
BD$+$46~3471 & V1578~Cyg    & A1 & 2.84$^{+0.07}_{-0.08}$ & 5.9$^{+0.6}_{-0.5}$ & 9.7$^{+1.9}_{-1.9}$   & ... & $199 \pm 11$ & ... &... & $-3 \pm 9$ &  O \\ 
BD$+$61~154  & V594~Cyg     & B8 & 1.95$^{+34}_{-24}$ & 3.41$^{+0.38}_{-0.38}$ & 2.42$^{+0.35}_{-0.35}$  & ... & $112 \pm 24$ & ... &... & $-16 \pm 18$ &  PC  \\ 
HD~17081     & $\pi$~Cet    & B8 & 2.750$^{+0.022}_{-0.022}$ & 4.65$^{+0.08}_{-0.08}$ & 4.84$^{+0.19}_{-0.19}$  & ... & $19.9 \pm 0.9$ & $0.76^{el}$ &  30.6 & [11.0,12.7] & O \\ 
HD~31293     & AB~Aur       & A0 & 1.76$^{+0.12}_{-0.11}$ & 2.50$^{+0.29}_{-0.13}$ & 2.62$^{+0.44}_{-0.44}$  & -6.85$^c$ & $116 \pm 6$ & 0.91$^{rl}$ & 279.8 & $24.7 \pm 4.7$ &  PC  \\ 
HD~31648     & MWC~480      & A4 & 1.18$^{+0.18}_{-0.15}$ & 1.93$^{+0.09}_{-0.14}$ & 1.93$^{+0.32}_{-0.32}$  & $<-7.23^c$ & $97.5 \pm 4.7$ & $0.78^{el}$ & 155.8 & $12.9 \pm 3.5$ &  PC  \\ 
HD~34282     &              & A3 & 1.13$^{+0.27}_{-0.22}$ & 1.59$^{+0.30}_{-0.07}$ & 1.66$^{+0.62}_{-0.37}$  & $-7.69^{+0.28}_{-0.59}$$^a$ & $105 \pm 6$ & $0.41^{el}$ & 115.1 & $16.2 \pm 4.8$ &  IPC  \\ 
HD~35187     &              & A2 & 1.15$^{+0.27}_{-0.20}$ & 1.93$^{+0.28}_{-0.04}$ & 1.58$^{+0.02}_{-0.02}$  & $-7.60^c$ & $93.3 \pm 2.8$ & ... &... & $27.0 \pm 2.1$ & O  \\ 
HD~35929     &              & F1 & 2.12$^{+0.07}_{-0.07}$ & 4.13$^{+0.23}_{-0.24}$ & 8.1$^{+0.7}_{-0.7}$   & $-6.37^{+0.18}_{-0.26}$ & $61.8 \pm 2.2$ & $0.85^{el}$ & 117.3 & $ 21.1 \pm 1.8$ &  O  \\ 
HD~36112     & MWC~758      & A5 & 1.81$^{+0.25}_{-0.19}$ & 2.90$^{+0.67}_{-0.43}$ & 4.4$^{+0.9}_{-0.9}$   & $-6.05^c$ & $54.1 \pm 4.9$ & $0.66^{el}$ &  72.0 & $17.8 \pm 3.7$ &  PC  \\ 
HD~36910     & CQ~Tau       & F2 & 1.69$^{+0.20}_{-0.16}$ & 2.93$^{+0.54}_{-0.37}$ & 5.1$^{+0.9}_{-0.9}$   & $<-8.30^c$ & $98 \pm 5$ & ... &... & $35.7 \pm 4.5$ &  IPC  \\ 
HD~36917     & V372~Ori     & B9$^*$ & 2.39$^{+0.07}_{-0.07}$ & 3.98$^{+0.25}_{-0.24}$ & 5.2$^{+0.6}_{-0.6}$   & ... & $127.1 \pm 4.6$ & $0.01^{el}$ & 127.1 & $26.3 \pm 3.6$ &  O \\ 
HD~36982     & LP~Ori       & B1.5$*$ & 3.22$^{+0.07}_{-0.07}$ & 6.70$^{+0.64}_{-0.37}$ & 3.42$^{+0.30}_{-0.30}$ & ... & $88 \pm 8$ & ... &... & $30 \pm 6$ &  O \\ 
HD~37258     & V586~Ori     & A1 & 1.44$^{+0.07}_{-0.07}$ & 2.28$^{+0.15}_{-0.16}$ & 1.94$^{+0.24}_{-0.24}$  & $-6.98^{+0.14}_{-0.17}$$^a$ & $200 \pm 14$ & $0.41^{el}$ & 219.2 & $31 \pm 12$ &  IPC \\ 
HD~37357     &              & A1 & 1.72$^{+0.07}_{-0.07}$ & 2.47$^{+0.13}_{-0.11}$ & 2.83$^{+0.35}_{-0.35}$  & $-6.42^{+0.09}_{-0.06}$$^a$ & ... &... & $124 \pm 7$ & $21.4 \pm 4.7$ &  O \\ 
HD~37806     & MWC~120      & B9 & 2.45$^{+0.07}_{-0.07}$ & 3.94$^{+0.23}_{-0.23}$ & 4.6$^{+0.5}_{-0.5}$   & $-6.85^c$ & $120 \pm 27$ & $0.75^{rl}$ & 181.4 & $47 \pm 21$ &  IPC  \\ 
HD~38120     &              & B9 & 1.62$^{+0.07}_{-0.07}$ & 2.49$^{+0.09}_{-0.09}$ & 1.91$^{+0.11}_{-0.11}$  & $-6.90^c$ & $97 \pm 17$ & ... &... & $28 \pm 12$ &  O  \\ 
HD~38238     & V351~Ori     & A6 & 1.79$^{+0.07}_{-0.07}$ & 2.88$^{+0.18}_{-0.18}$ & 4.38$^{+0.44}_{-0.44}$  & ... & $99.8 \pm 4.2$ & ... &... & $15.0 \pm 2.9$ &  IPC  \\ 
HD~50083     & V742~Mon     & B4 & 4.15$^{+0.12}_{-0.12}$ & 12.1$^{+1.1}_{-1.1}$ & 10.0$^{+1.0}_{-1.0}$   & ... & $233 \pm 22$ & ... &... & $-0.4 \pm 1.2$ &  O  \\
HD~52721     & GU~CMa       & B3 & 3.77$^{+0.35}_{-0.31}$ & 9.1$^{+2.4}_{-1.4}$ & 5.0$^{+1.2}_{-1.2}$   & $-5.00^{+0.23}_{-0.13}$$^a$ & $215 \pm 18$ & ... &... & $21 \pm 14$ &  O  \\ 
HD~53367     & MWC~166      & B1 & 4.50$^{+0.25}_{-0.20}$ & 16.1$^{+2.7}_{-1.6}$ & 7.1$^{+1.6}_{-1.6}$   & $<-8.92^c$ & $41 \pm 7$ & ... &... & $47.2 \pm 4.8$ &  O \\ 
HD 68695     &              & F2$^*$ & 1.80$^{+0.14}_{-0.17}$ & 2.64$^{+0.31}_{-0.30}$ & 3.3$^{+0.6}_{-0.6}$   & $-7.78^{+0.30}_{-0.38}$$^a$ & $43.8 \pm 2.6$ & ... &... & $20.3 \pm 1.7$ &  PC \\ 
HD~72106     &              & A0$^*$ & 1.34$^{+0.28}_{-0.26}$ & 2.40$^{+0.3}_{-0.3}$ & 1.3$^{+0.5}_{-0.5}$   & $-6.21^{+0.15}_{-0.18}$$^a$ & $41.0 \pm 0.3l$ & ... &... & $22 \pm 1$ &  O \\ 
HD72106\_B   &              &     & 0.96$^{+0.27}_{-0.27}$ & 1.9$^{+0.2}_{-0.2}$ & 1.3$^{+0.5}_{-0.5}$   & ... & $53.9 \pm 1.0l$ & ... &... & $22 \pm 1$ & \\
HD~76534     &              & B2$^*$ & 3.75$^{+0.08}_{-0.08}$ & 9.0$^{+0.6}_{-0.6}$ & 7.7$^{+1.6}_{-1.6}$   & $\leq -6.95^a$ & $68 \pm 30$ & ... &... & $23 \pm 18$ &  O \\ 
HD~98922$^a$ &              & B9$^*$ & 2.48$^{+0.15}_{-0.15}$ & 4.0$^{+0.5}_{-0.5}$ & $5.2^{+0.3}_{-0.3}$ & $\leq -6.97$   & $50.0 \pm 3.0$ & ... &... & $0.2 \pm 2.2$ &  PC \\ 
HD~101412 &             & B9/A0$^*$ & 1.36$^{+0.23}_{-0.23}$ & 2.0$^{+0.1}_{-0.1}$ & $1.7^{+0.1}_{-0.1}$  & $\leq -7.61^a$ & $3 \pm 1^d$ & $0.17^e$ &  3.0 & $\sim 16.5 \pm 0.5^f$ & IPC \\ 
HD~114981    & V958~Cen     & B3/5$^*$ & 3.56$^{+0.34}_{-0.24}$ & 7.9$^{+2.4}_{-1.3}$ & 7.0	2.0	-2.0   & $\leq -5.48^a$ & $239 \pm 13$ & ... &... & $-50 \pm 11$ &  O  \\ 
HD 139614    &              & F0$^*$ & 1.10$^{+0.15}_{-0.18}$ & 1.76$^{+0.15}_{-0.08}$ & 2.06$^{+0.42}_{-0.42}$  & $-7.63^{+0.20}_{-0.30}$$^a$ & $24.1 \pm 3.0$ & $0.85^{rl}$ &  45.7 & $0.3 \pm 2.3$ &  O \\
HD~141569    &              & A0 & 1.49$^{+0.06}_{-0.06}$ & 2.33$^{+0.20}_{-0.12}$ & 1.94$^{+0.21}_{-0.21}$  & $-7.65^{+0.33}_{-0.47}$$^a$ & $228 \pm 10$ & ... &... & $-12 \pm 7$ &  O  \\ 
HD~142666    & V1026~Sco    & A5 & 1.44$^{+0.11}_{-0.13}$ & 2.15$^{+0.20}_{-0.19}$ & 2.82$^{+0.41}_{-0.41}$  & $\leq -8.38^a$ & $5.3 \pm 3.1$ & $0.50^{rl}$ &   6.1 & $-7.0 \pm 2.7$ &  IPC  \\ 
HD~144432    &              & A7 & 1.28$^{+0.11}_{-0.13}$ & 1.95$^{+0.18}_{-0.16}$ & 2.59$^{+0.40}_{-0.40}$  & $-7.38^{+0.22}_{-0.40}$$^a$ & $78.8 \pm 4.2$ & $0.91^{rl}$ & 190.1 & $-3.0 \pm 3.5$ &  PC  \\ 
HD~144668    & HR~5999      & A7$^*$ & 1.56$^{+0.15}_{-0.18}$ & 2.31$^{+0.29}_{-0.28}$ & 3.0$^{+0.6}_{-0.6}$   & $-6.25^{+0.16}_{-0.19}$$^a$ & $199 \pm 11$ & $0.61^{rl}$ & 251.1 & $-10 \pm 8$ &  O \\ 
HD~145718    & V718~Sco     & A4 & 1.29$^{+0.11}_{-0.13}$ & 1.93$^{+0.14}_{-0.08}$ & 2.25$^{+0.33}_{-0.33}$  & $\leq -8.51$ & $113.4 \pm 4$ & $0.47^{rl}$ & 128.5 & $-3.6 \pm 2.3$ &  O  \\ 
HD~150193    & MWC~863      & A1 & 1.79$^{+0.11}_{-0.13}$ & 2.56$^{+0.22}_{-0.19}$ & 2.89$^{+0.48}_{-0.48}$  & $-7.45^{+0.25}_{-0.43}$$^a$ & $108 \pm 5$ & $0.85^{rl}$ & 205.0 & $-4.9 \pm 3.9$ &  PC  \\ 
HD~152404\_A & AK~Sco       & F5$^*$ & 0.95$^{+0.21}_{-0.21}$ & 1.66$^{+0.29}_{-0.21}$ & 2.4$^{+0.5}_{-0.5}$   & $\leq -7.90$$^a$ & $18.2 \pm 1.7$ & ... &... & $-17.0 \pm 1.3$ &  PC \\
HD~152404\_B & AK\_Sco\_B   &   & 0.71$^{+0.21}_{-0.21}$ & 1.43$^{+0.20}_{-0.09}$ & 1.79$^{+0.38}_{-0.38}$  & ... & $17.6 \pm 0.9$ & ... &... & $14.3 \pm 0.9$ & \\
HD~163296    &              & A1 & 1.52$^{+0.08}_{-0.08}$ & 2.23$^{+0.22}_{-0.07}$ & 2.28$^{+0.23}_{-0.23}$  & $-7.49^{+0.14}_{-0.30}$$^a$ & $129 \pm 8$ & $0.67^{rl}$ & 173.8 & $-9 \pm 6$ &  PC  \\ 
HD~169142    &              & A7 & 0.88$^{+0.21}_{-0.28}$ & 1.69$^{+0.06}_{-0.14}$ & 1.64$^{+0.20}_{-0.20}$  & $-7.40^g$ & $47.8 \pm 2.3$ & $0.92^{rl}$ & 122.0 & $-0.4 \pm 2.0$ &  O \\ 
HD~174571    & MWC~610      & B3 & 3.58$^{+0.21}_{-0.21}$ & 8.0$^{+1.2}_{-1.0}$ & 4.7$^{+0.6}_{-0.6}$   & ... & $219 \pm 31$ & ... &... & $14 \pm 24$ &  O  \\ 
HD~176386    &              & B9$^*$ & 1.91$^{+0.09}_{-0.09}$ & 3.02$^{+0.23}_{-0.26}$ & 2.28$^{+0.24}_{-0.24}$  & $-8.11^g$ & $175 \pm 6$ & ... &... & $-2 \pm 5$ &  O \\ 
HD~179218    & MWC~614      & A0 & 2.26$^{+0.14}_{-0.12}$ & 3.66$^{+0.44}_{-0.34}$ & 4.8$^{+0.7}_{-0.7}$   & $-6.59^c$ & $68.8 \pm 2.9$ & $0.66^{el}$ &  91.6 & $15.1 \pm 2.3$ &  O  \\ 
HD~190073    & V1295~Aql    & A1 & 1.92$^{+0.12}_{-0.12}$ & 2.85$^{+0.25}_{-0.25}$ & 3.60$^{+0.5}_{-0.5}$   & $-5.00 \pm 0.25^b$ & [0-8.3] & $0.86^{rl}$ & 0-16.3 & $0.21 \pm 0.10$ & PC  \\ 
HD~200775~A  & MWC~361~A    & B2$^*$ & 3.95$^{+0.30}_{-0.30}$ & 10.7$^{+2.5}_{-2.5}$ & 10.4$^{+4.9}_{-4.9}$   & ... & $26 \pm 2$ & ... &... & [-23.3,8.2] &  IPC  \\ 
HD~200775~B  & MWC~361~B    & ? & 3.77$^{+0.30}_{-0.30}$ & 9.3$^{+2.1}_{-2.1}$ & 8.3$^{+3.9}_{-3.9}$  & ... & $59 \pm 5$ & ... &... & [-21.1,9.3] & \\
HD~216629~A  & IL~Cep~A     & B4 & 2.58$^{+0.20}_{-0.20}$ & ... & ... & ... & $179 \pm 27$ & ... &... & [-39,31] &  O  \\ 
HD~216629~B  & IL~Cep~B     & ? & ... & ... & ? & $152 \pm 33$ & ... & ... & ... & [-87,-30] & \\
HD~244314    & V1409~Ori    & A1 & 1.45$^{+0.07}_{-0.07}$ & 2.33$^{+0.08}_{-0.23}$ & 2.07$^{+0.26}_{-0.26}$  & $-7.12^{+0.20}_{-0.25}$ & $51.9 \pm 2.2$ & ... &... & $22.5 \pm 1.8$ &  PC \\ 
HD~244604    & V1410~Ori    & A4 & 1.74$^{+0.07}_{-0.07}$ & 2.66$^{+0.15}_{-0.15}$ & 3.69$^{+0.34}_{-0.34}$  & $-7.22^{+0.26}_{-0.32}$$^a$ & $98.3 \pm 1.8$ & $0.58^{el}$ & 120.7 & $26.8 \pm 1.6$ &  PC  \\ 
HD~245185    & V1271~Ori    & A1 & 1.40$^{+0.09}_{-0.10}$ & 2.19$^{+0.27}_{-0.12}$ & 1.85$^{+0.20}_{-0.20}$  & $\leq -7.29^a$ & $118 \pm 22$ & ... &... & $16 \pm 16$ &  O \\ 
HD~249879    &              & A2 & 2.31$^{+0.19}_{-0.25}$ & 4.0$^{+0.8}_{-0.8}$ & 5.9$^{+1.8}_{-1.8}$   & -8.00 & $254 \pm 26$ & ... &... & $11 \pm 20$ &  O \\ 
HD~250550    & V1307~Mon    & B8 & ... & ... & ... & $-5.63^{+0.14}_{-0.11}$$^a$ & $79 \pm 9$ & $0.62^{el}$ & 100.7 & $-22 \pm 8$ &  PC  \\ 
HD~259431    & MWC~147      & B6$^*$ & 3.35$^{+0.12}_{-0.14}$ & 7.1$^{+0.8}_{-0.8}$ & 8.0$^{+1.6}_{-1.6}$   & $-6.24^{+0.19}_{-0.20}$$^h$ & $83 \pm 11$ & $0.98^{rl}$ & 417.1 & $26 \pm 8$ &  PC \\ 
HD~275877    & XY~Per & A2  & 1.21$^{+0.47}_{-0.30}$ & 1.95$^{+0.46}_{-0.09}$ & 1.65$^{+0.11}_{-0.11}$    & $-7.02^c$ & $224 \pm 12$ & ... &... & $2 \pm 10$ &  IPC  \\ 
HD~278937    & IP~Per & A3  & 1.21$^{+0.08}_{-0.09}$ & 1.86$^{+0.10}_{-0.06}$ & 2.10$^{+0.23}_{-0.23}$    & ... & $79.8 \pm 2.9$ & ... &... & $13.7 \pm 2.1$ &  PC  \\ 
HD~287823~A  &              & A0 & 1.79 & 2.5 & 2.6 & $-7.13^{+0.18}_{-0.23}$$^a$ & $10.3 \pm 1.5$ & ... &... & $-0.3 \pm 1.1$ &  O  \\ 
HD~287823~B  &              & ? & 0.82 & 1.6 & 1.8 & ? & $8.2 \pm 3.3$ & $54.0 \pm 1.6$ \\
HD~287841    & V346~Ori     & A7 & 1.05$^{+0.07}_{-0.07}$ & 1.72$^{+0.15}_{-0.05}$ & 1.96$^{+0.20}_{-0.20}$  & $\leq -7.82^a$ & $115.8 \pm 4.2$ & ... &... & $20.0 \pm 3.6$ &  IPC  \\ 
HD~290409    &              & A2 & 1.32$^{+0.06}_{-0.06}$ & 2.04$^{+0.18}_{-0.18}$ & 1.75$^{+0.20}_{-0.20}$  & $\leq -7.31^a$ & $250 \pm 120$ & ... &... & $1 \pm 70$ &  O \\ 
HD~290500    &              & A2 & 1.22$^{+0.07}_{-0.07}$ & 1.96$^{+0.21}_{-0.06}$ & 1.68$^{+0.09}_{-0.09}$  & $-6.11^{+0.17}_{-0.17}$ & $85 \pm 15$ & ... &... & $29 \pm 11$ &  IPC \\ 
HD~290770    &              & B9 & 1.91$^{+0.07}_{-0.07}$ & 2.86$^{+0.27}_{-0.21}$ & 2.49$^{+0.44}_{-0.44}$  & $-6.74^{+0.12}_{-0.14}$ & $240 \pm 100$ & ... &... & $4 \pm 60$ &  PC \\ 
HD~293782    & UX~Ori       & A1 & 2.98$^{+0.07}_{-0.07}$ & 6.72$^{+0.42}_{-0.43}$ & 12.1$^{+1.5}_{-1.5}$   & $\leq -7.26^a$ & $221 \pm 13$ & $0.87^{el}$ & 448.2 & $12 \pm 10$ &  IPC  \\ 
             & MWC~1080     & B1 & 5.77$^{+0.20}_{-0.26}$ & 17.4 & 7.3 & ... &  ... & ... & ... &... &  PC  \\ 
             & VV~Ser       & B7 & 2.51$^{+0.28}_{-0.42}$ & 4.0$^{+0.8}_{-0.8}$ & 3.1$^{+0.9}_{-0.9}$ & -7.50 & $124 \pm 24$ & $0.60^{rl}$ & 155.0 & $51 \pm 18$ &  IPC \\ 
             & LkHa~215~A   & B7 & 3.08$^{+0.10}_{-0.10}$ & 5.8 & 5.9 & $-6.72^{+0.23}_{-0.17}$$^h$ & $210 \pm 70$ & ... &... & $0 \pm 40$ &  O  \\ 
             & LkHa~215~B   & ? & 3.08$^{+0.10}_{-0.10}$ & 5.8 & 5.9 & ...  & $11.7 \pm 4.6$ & ... &... & [12,22] & \\
\\
\\
\hline
\multicolumn{11}{l}{PC = P-Cygni profile (blueshifted absorption); IPC = inverse P-Cygni profile (redshifted absorption), O = other} \\ 
\multicolumn{11}{l}{$^*$ spectral type from Simbad \citep{wen00}, $^a$ \citet{fai15}, $^b$ \citet{men11}, $^c$ \citet{cau14}, $^d$ \citet{cow10}, } \\
\multicolumn{11}{l}{$^e$ \citet{fed08}, $^f$ \citet{hub09}, $^g$ \citet{gar06}, $^h$ \citet{don11}} \\
\multicolumn{11}{l}{$^{rl}$ ring model from \citet{laz17}, $^{el}$ ellipsoidal model from \citet{laz17}} \\
\end{longtable*}
\end{center}
\end{footnotesize}
\end{longrotatetable}

\end{document}